\documentclass[reqno,11pt]{amsart}
\usepackage[utf8]{inputenc}
\usepackage{xcolor}
\usepackage{enumitem}
\usepackage[linktocpage]{hyperref}
\usepackage{graphicx}
\usepackage{amscd}
\usepackage{slashed}
\usepackage{amssymb}
\usepackage{mathtools} 
\usepackage{pstricks}
\usepackage[mathscr]{eucal}
\usepackage{soul}
\usepackage{comment}
\numberwithin{equation}{section}
\allowdisplaybreaks[1]
\definecolor{labelkey}{gray}{.65}

\title[CFS, NCG and GTD]{Causal Fermion Systems, Non-Commutative Geometry and Generalized Trace Dynamics}

\author[S.\ Farnsworth]{Shane Farnsworth}
\address{Fakult\"at f\"ur Mathematik \\ Universit\"at Regensburg \\ D-93040 Regensburg \\ Germany \& Guest position at Max-Planck Institute for Gravitational Physics \\Am Mühlenberg 1, 14476 Potsdam, Germany}
\email{Shane.Farnsworth@aei.mpg.de}

\author[F.\ Finster]{Felix Finster}
\address{Fakult\"at f\"ur Mathematik \\ Universit\"at Regensburg \\ D-93040 Regensburg \\ Germany}%
\email{finster@ur.de}

\author[C.F.\ Paganini]{Claudio F. Paganini}
\address{Fakult\"at f\"ur Mathematik \\ Universit\"at Regensburg \\ D-93040 Regensburg \\ Germany}%
\email{claudio.paganini@ur.de}

\author[T.\ Singh]{Tejinder P. Singh \\ \\ March 2026}
\address{Inter-University Centre for Astronomy and Astrophysics,
Post Bag 4, Ganeshkhind, Pune 411007, India and
Tata Institute of Fundamental Research, Homi Bhabha Road, Mumbai 400005, India}
\email{tejinder.singh@iucaa.in; tpsingh@tifr.res.in}

\newtheorem{Def}{Definition}[section]

\newtheorem{Lemma}[Def]{Lemma}

\newcommand{\Thanks}{\vspace*{.5em} \noindent \thanks}
\newcommand{\beq}{\begin{equation}}
\newcommand{\eeq}{\end{equation}}
\newcommand{\Proof}{\begin{proof}}
\newcommand{\QED}{\end{proof} \noindent}

\newcommand{\la}{\langle}
\newcommand{\ra}{\rangle}

\newcommand{\Sl}{\mathopen{\prec}}
\newcommand{\Sr}{\mathclose{\succ}}

\newcommand{\C}{\mathbb{C}}
\newcommand{\R}{\mathbb{R}}
\newcommand{\1}{\mbox{\rm 1 \hspace{-1.05 em} 1}}

\newcommand{\N}{\mathbb{N}}

\renewcommand{\H}{\mathscr{H}}

\newcommand{\SL}{{\rm{SL}}}

\newcommand{\bep}{\begin{pmatrix}}
\newcommand{\enp}{\end{pmatrix}}

\newcommand{\F}{{\mathscr{F}}}

\renewcommand{\L}{{\mathcal{L}}}
\newcommand{\Sact}{{\mathcal{S}}}

\newcommand{\Lin}{\text{\rm{L}}}

\newcommand{\T}{{\mathscr{T}}}

\newcommand{\scrM}{\myscr M}

\newcommand{\lec}{\text{\rm{le}}}
\newcommand{\hec}{\text{\rm{he}}}
\newcommand{\Sig}{\mathscr{S}}

\DeclareFontFamily{OT1}{rsfso}{}
\DeclareFontShape{OT1}{rsfso}{m}{n}{ <-7> rsfso5 <7-10> rsfso7 <10-> rsfso10}{}
\DeclareMathAlphabet{\myscr}{OT1}{rsfso}{m}{n}

\DeclareMathOperator{\tr}{Tr}

\DeclareMathOperator{\diag}{diag}

\DeclareMathOperator{\supp}{supp}

\newcommand{\bitem}{\begin{itemize}[leftmargin=2.5em]}
\newcommand{\eitem}{\end{itemize}}

\newcommand{\fermi}{{\mathrm{f}}}

\begin{document}

\begin{abstract}
We compare the structures and methods in the theory of causal fermion systems
with generalized trace dynamics
and non-commutative geometry. Although the three theories differ on many aspects, they agree in that the geometric structure to be recovered in the continuum limit is not the bare spacetime but a suitable fiber bundle.
Furthermore, the comparison leads us to the conclusion that the key innovation in causal fermion systems lies in the manner in which the relation between different spacetime points is encoded. The role of Synge's classical world function $\sigma(x,y)$ that encodes the geodesic distance between any two points in the manifold, is taken by a generalized two-point correlator. We show that this idea can be transferred to non-commutative geometry and generalized trace dynamics. 
\end{abstract}

\maketitle

\tableofcontents

\section{Introduction} \label{secintro}

Physical phenomena can  often be  modelled effectively  in  several different mathematical frameworks. The archetypal example is that of a fluid, which at one level can be modelled using macroscopic variables like temperature and viscosity, while at a deeper level macroscopic fluid behavior is seen to emerge from the dynamics of statistical ensembles of molecules. More modern examples include dualities such as the AdS/CFT correspondence~\cite{Maldacena98}, or the loop equation description of turbulent flow~\cite{Migdal24}, which hold in appropriate limits. It remains an open question, which shift in perspective and description will turn out to be most useful for quantizing gravity. There are already a proliferation of proposals, and any approach that has a chance of succeeding must make contact with our current models of physics written in terms of effective field theory. In addition, the class of all such approaches must also cohere with one another at least within testable limits. An obvious question is whether the coherence between competing approaches can be leveraged to single out potential new dualities, or promising new avenues towards a more complete description of nature. 

 This paper explores and compares three approaches to unifying gravity with the Standard-Model of particle physics: Finster's causal fermion systems (CFS), Connes' Non-Commutative Geometry (NCG)~\cite{connes}, and Adler's Trace Dynamics (TD) as well as its generalization to include gravitation (GTD).
These three approaches have been selected because, while they differ significantly from one another, they also host a number of commonalities that highlight fundamental features of the underlying physics being modelled. Each approach uses different fundamental mathematical objects to describe known physics, offering surprising reinterpretations of well-known phenomena and suggesting new paths forward for theoretical development. Our goal is to provide a clear and concise overview of these approaches and how they compare and contrast from one another. The aim is to facilitate the porting of lessons learned between approaches, to draw connections between fields, and to sever dead ends. We highlight the key tools that these approaches use and mention open problems within each framework. Moreover, we point out areas in which they offer more explanatory power than the usual approach of effective field theory, and discuss potential applications beyond their original contexts. Our comparison aims to facilitate communication between CFS, NCG, and GTD. However, it should not be necessary to have any background in these approaches to understand or benefit from this paper.

This article is part of a series of papers that compare the structures and ideas of different approaches to fundamental physics that was started with~\cite{ethcfs, mmt-cfs, oct-cfs} and will be continued in~\cite{cfstdqg}. Each of these papers contains an introduction to the theories under consideration, including an overview of their accomplishments. These overviews serve as a starting point for the reader to get familiar with the approaches. The papers proceed with a detailed comparison of these theories under consideration. Ultimately, the goal is to motivate the community to establish an extensive collection of such articles as a sort of ``Rosetta stone'' for approaches to fundamental physics. The hope is that such a set of dictionaries will facilitate the exchange of ideas across approaches and thereby catalyze progress in the foundations of physics.

\subsection*{Organization of the Paper}
In Section~\ref{secoverview} we give an overview of the three theories under consideration, starting with NCG in Section~\ref{secoverview_NCG}, moving on to CFS in Section~\ref{secoverview_CFS} and ending with GTD in Section~\ref{secoverview_TD}. Each of these sections covers both the fundamental structures of the theories as well as the principles that govern the dynamics of physical systems according to these theories. 
The core of the paper is Section~\ref{seccompare} where we compare and contrast the three theories, starting from their fundamental structures in Section~\ref{secfundamental}. In Section~\ref{secemergence} we discuss how the classical continuum spacetime emerges in each of the theories. In Section~\ref{secquantization} we discuss in what sense each of the theories can be considered to be ``quantum''. In Section~\ref{secaction} we compare the action principles of the different theories and in Section~\ref{secconserv} we investigate how the unitary invariance of each gives rise to conservation laws. Transitioning towards the phenomenology of the theories, in Section~\ref{seccollapse} we explore how the three theories relate to collapse models. We then proceed with a discussion of the derivation of the Standard Model of particle physics in each of the approaches. In this context, we show that the key ideas from CFS on constructing the spacetime from correlation geometry can be adapted to both NCG and GDT, while NCG can serve as a constraint for the construction of different matter models in CFS. 

\subsection*{Author Contributions} The sections specific to one theory were written by the respective domain experts, Felix Finster and Claudio F. Paganini for Causal Fermion Systems (Section \ref{secoverview_CFS}), Shane Farnsworth for Non-Commutative Geometry (Section \ref{secoverview_NCG}) and Tejinder P. Singh for Generalized Trace Dynamics (Section \ref{secoverview_TD}). This also applies to the comparison Section \ref{seccompare} where however, the discussion sections were written in collaboration. Furthermore, all authors contributed to the overall structure of the manuscript otherwise not specified here.

\section{Overview of NCG, CFS, and GTD} \label{secoverview}

In this section, we provide a succinct overview of NCG, CFS, and GTD as independent theories. We break each discussion into two parts: First, we describe the \emph{basic objects} that are used to model physical phenomena in each formalism. In the more familiar language of effective field theory, space-time is modelled as a pseudo-Riemannian manifold, with physical content, including the space-time metric, gauge bosons, scalars, and fermions, represented as fields on this manifold. We will describe how NCG, CFS, and GTD reformulate these fundamental concepts in terms of mathematical structures that shift focus away from conventional spacetime descriptions. We then explain the benefit of each reformulation and how one recovers the usual picture. Secondly, we describe the way in which the \emph{dynamics} is encoded in each framework. In the standard approach, one makes use of the standard model action coupled to the Einstein-Hilbert action, or something similar. Here we describe the use of the spectral action for both NCG and GTD, and the causal action for CFS.

\subsection{NCG Overview} \label{secoverview_NCG} The NCG approach to physics~\cite{Connes2008kx} is to unify gravity with the standard model of particle physics by searching for a geometric description of the standard model, which is similar to the geometric description we already have for gravity. The geometry is not assumed to be Riemannian, but is instead allowed to be a more general kind of `non-commutative' space, which is directly chosen to match experiment, and which is formulated in terms of an operator algebra and a Dirac operator represented on a Hilbert space. Such `spectral triples' of algebraic objects can be used to reconstruct the usual manifold and metric data of Riemannian geometries~\cite{Connes2013SF}, but they also readily generalize to reconstruct more general particle content living on a manifold. In the most conservative setting the NCG approach unifies standard model physics with gravity as a classical theory~\cite{chamseddine-connes-marcolli}, and provides novel geometric re-interpretations for many of the elements of classical field theory. However, in less conservative approaches, the framework allows one to replace the base manifold with non-commutative analogs, thereby providing a novel setting for exploring the quantization of spacetime itself~\cite{Walter22, Chamseddine_2014}.

\subsubsection{NCG Overview: Fundamental Elements}

NCG provides a flexible framework for modelling physical systems geometrically. The framework is general enough to describe commutative Riemannian geometries\cite{Connes2013SF}, as well as broad classes of abstract finite-dimensional, `non-commutative', and discrete spaces~\cite{Landi97}. This freedom allows one to ask in an agnostic manner: `What kind of geometry best captures the observed physics that I am trying to model?'. A prime example showcasing the advantages of remaining agnostic in this way is the comparison between Kaluza-Klein theory~\cite{KKtheory} (KK theory) and Connes' non-commutative standard model~\cite{connes}. 
Kaluza-Klein theory assumes the geometry that describes electromagnetism coupled to
Einstein-Hilbert gravity to be pseudo-Riemannian with a compactified internal space. In doing so, the dynamics and infinite-dimensional symmetries of gravitational
theories are appropriated for the description of the predominantly static, finite-dimensional, internal spaces of gauge theories. This leads to instabilities
within the compactified internal space~\cite{wit82}, and the $5$-dimensional metric also gives rise to an unwanted dilaton field, in addition to the desired $4$-dimensional metric and a $U(1)$ gauge field. In contrast, while challenges remain with the non-commutative standard model~\cite{chamseddine-connes-marcolli}, it does not suffer from these specific problems.

The NCG framework provides modelling flexibility by replacing the familiar metric and manifold data used for describing Riemannian geometries with algebraic structures that capture the same kinds of metric, differential, and topological data, but which are more readily generalizable. Specifically, the fundamental elements of a non-commutative geometry are described by a spectral triple~\cite{connes} denoted by
\begin{equation} T= (A,\H,D) \:, \end{equation}
which consists of a `coordinate' algebra $A$, a Dirac operator $D$, and a Hilbert space $\H$, on which $A$ and $D$ are represented.  At first sight, spectral triples appear rather alien to a physicist, and certainly not very geometric, but the key ideas behind their construction are, in fact, very intuitive and stem from two simple, but deep and beautiful observations that we now review. 

The first observation has to do with the deep relationship between operator algebras and topological spaces, and hence the reason for the elements $(A,\H)$ in a spectral triple. An illustrative example of the correspondence between algebras and topological spaces can be seen when considering algebras of complex $n\times n$ matrices $A = M_n(\mathbb{C})$ represented on complex Hilbert spaces $\H=\mathbb{C}^n$. The pure states on such algebras $\rho_\psi(a) = \langle \psi|a|\psi\rangle$, $a\in \text{End}(H)$, are maps to the complex numbers expressed in the usual way in terms of normalized elements $\psi\in \H$. The possible choices of $\psi$ map out a compact manifold of states. In the case $n=2$, the space of pure states is given by the well-known Bloch sphere~\cite{Schultz1998,Farnsworth_2020},
providing a tangible and explicit example where one is able to see the correspondence between an algebra on the one hand and a topological space derived from the representation of the algebra on a Hilbert space on the other.  Beyond the finite-dimensional case, another well-known equivalence~\cite{Gelfand1943} also exists between commutative $C^\ast$-algebras and locally compact Hausdorff spaces, where once again the pure states on the algebra play the role of points. NCG~\cite{connes,Connes2008kx} takes the correspondence between algebras and topologies to the logical limit, completely replacing the usual geometric descriptions of manifolds, instead encoding differential topological data with algebras. The rich variety of available algebras and their representations on Hilbert spaces then gives rise to broad classes of topological spaces that go beyond the usual Riemannian paradigm, including non-commutative~\cite{Connes_1998},  fractal~\cite{mar-ball} and discrete \cite{Iochum_2001} spaces~\cite{connes-reality}. More recently the formalism has also been extended  to describe non-associative spaces~\cite{Besnard_2022,farnsworth2025npointexceptionaluniverse,farnsworth2025spectralgeometryexceptionalsymmetry, boyle-farnsworth, Boyle_2020, Farnsworth2013NonassociativeGA,Farnsworth_2020,Farnsworth_2015}.

Beyond topology, the second key idea behind NCG concerns the relationship between derivative operators and metric data, which relates specifically to the role of the Dirac operator $D$ in a spectral triple. On a smooth manifold $M$, the usual definition of distance $d(x,y)$ between any two points $x$ and $y$
is given by the length of the shortest path connecting them. 
 However, there exists an alternative formulation of distance that does not rely on geodesics and is thus far more adaptable to generalized, algebraic spaces where a conventional notion of geodesics may no longer apply. On any Riemannian manifold, one can always find a function $f$ such that $|f(x) -f(y)| = d(x,y)$; this is possible because any function with finite change between two points can be rescaled to match the desired distance. The challenge lies in finding such a function without first knowing $d(x,y)$. The solution is to restrict to smooth functions whose gradients are bounded in norm by 1, and then search for a function that maximizes $|f(x) -f(y)|$. Remarkably, this procedure recovers the standard geodesic distance of Riemannian geometry~\cite{Connes2008kx}. For this construction to work, the only additional structure required, beyond the algebra
$A$ of smooth functions, is a first-order differential operator that encodes gradient information. In non-commutative geometry, this role is played by the Dirac operator $D$, which enables a generalized notion of distance~\cite{Connes1989,Krajewski2001} even in settings where the smooth manifold structure is absent. This includes non-commutative, finite-dimensional, or discrete spaces, where a suitable coordinate algebra still encodes the topological data of the space~\cite{Krajewski2001}.

The above discussion should provide a conceptual picture as to why and how the elements in a spectral triple $T= (A,\H,D)$ encode a geometry.  When the spectral framework is used to model particle physics,  the `internal' space of the NCG that describes the standard model is given by a finite-dimensional,  discrete, and mainly non-dynamical geometry. It is the `non-dynamical' character of this geometry that allows one to circumvent many of the problems that arise in geometric descriptions of physics that rely on the compactification of smooth manifolds (e.g. string theory). To understand what `non-dynamical' means exactly, consider again the Bloch sphere, which is described by the pure states of the algebra of $2\times 2$ complex matrices represented on $\H=\mathbb{C}^2$. The metric on the Bloch sphere is essentially `frozen' because it is inherited from the embedding of the Bloch sphere as a projective space in $\mathbb{C}^2$. This means that the metric is fixed, rather than being subject to dynamical fluctuations as in conventional 2-dimensional Riemannian manifolds, where the metric enjoys an infinite-dimensional reparametrization invariance, which allows for a rich set of fluctuations and deformations. In the case of the Bloch sphere, the only non-trivial symmetry present is the finite-dimensional $\text{SU}(2)$ group of rotations. These rotations act by moving points around without altering the fixed metric structure and correspond to the automorphisms of the representation of the coordinate algebra. 

When constructing the internal spaces of gauge theories with multiple gauge sectors, one begins with a non-simple, finite-dimensional coordinate algebra $A=\oplus_{i=1}^n A^i$. From a geometric standpoint, the factors in the algebra $A^i$ correspond to discrete points, while from a physical standpoint the  symmetries of the representation of the $A_i$ correspond to the factors in the gauge group of the particle theory being modelled. While the geometric points themselves are non-dynamical, the distance between the points is, and is encoded within the Dirac operator.  Physically, this data corresponds to scalar fields that couple to the various factors of a gauge theory, just as the Higgs field couples to the left- and right-handed sectors of the standard model. One can immediately see the benefit of the NCG approach to the standard model: Rather than whittling down the infinite-dimensional diffeomorphism group of a Riemannian manifold to obtain the finite-dimensional gauge group of the standard model, one instead can select the factors $A_i$ and their representation to precisely match the gauge symmetries of the physics being modelled. Furthermore, rather than attempting to stabilize an infinitely malleable Riemannian manifold, one deals instead with fixed compact spaces, with the dynamics isolated to the metric data `between' the compact spaces, which corresponds precisely to Higgs-like fields which are observed in nature, as opposed to dilatons which are not. 

In practice, the most conservative approach to building spectral geometries corresponding to gauge theories is to take the product of a standard Riemannian geometry $T_c = (A_c, \H_c, D_c)$, which encodes a classical external spacetime, and a finite dimensional `internal' space $T_F = (A_F , \H_F , D_F)$  that  describes particle species and representations. In this case the external geometry is given exactly as one might expect: the algebra is nothing other than the smooth functions
over a 4-manifold, represented on square integrable Dirac spinors, and  the Dirac operator is the standard curved space Dirac operator:
\begin{align}
    A_c &= C^\infty(M), & \H_c &= L^\infty(M, S), &   D_c &= e_a^\mu \gamma^a \nabla_\mu^S
\end{align}
Of course, ultimately, one would like to describe the external spacetime itself by a truly non-commutative space, however, here we only review the construction of standard gauge theories coupled to Einstein-Hilbert gravity. 

For the internal space, one selects a finite-dimensional and discrete algebra $A_F = \oplus^i A_i$, along with its representation on a suitable complex vector space $\H_F = \mathbb{C}^n$, in order to reproduce the fermion representations and gauge symmetries of the physics being modelled. The finite Dirac operator $D_F\in \text{End}(\H_F)$ is nothing other than a matrix, subject to a number of geometric constraints. The product geometry  $ T = (A, \H, D)$ is given by~\cite{Farnsworth_2017}

\begin{align}\label{eq:NCGsplit}
A &= C^\infty(M,A_F), & \H &= \H_c\otimes \H_F, & D = D_c \otimes D_F + \gamma_5\otimes D_F,   
\end{align}
where $\gamma_5$ is Dirac's standard $\mathbb{Z}_2$ grading operator. Just as one replaces partial derivatives with gauge covariant derivatives in standard gauge theory, there is a standard procedure for `fluctuating' the Dirac operator to obtain a gauge covariant object $D \rightarrow  D + F$, where the connection term $F$ is given by
\begin{equation} F = e^\mu_a\gamma^a F_\mu(x)^i \otimes T_i + \gamma_5\otimes \Phi \:, \end{equation}
with real coefficients  $F_\mu^i(x)$ and where the $T_i$ form a basis which generates the gauge group. The first term on the left corresponds to gauge fields, and
the second term corresponds to Higgs fields $\Phi$. 

\subsubsection{NCG Overview: The Spectral Action Principle} \label{sec:spectralaction}
Having reformulated particle and gauge content in terms of spectral triples, a natural question emerges: How should dynamics be encoded within the setting of spectral triples and how can contact be made between NCG and the usual description
of physics given in terms of effective field theory. An answer to both questions is provided by the spectral action, which is a functional on the space of spectral triples~\cite{Chamseddine_1997}
\begin{equation} S(D) = \text{Tr}[f(D/\Lambda)] \:, \end{equation}
where $D$ is a generalized Dirac operator, and $\Lambda$ provides a scale at which the theory is defined.  This functional is well defined when $D$ has compact resolvent and when $f(x)$ is a smooth cutoff function that ensures convergence. It is valued at 1 for arguments less than 1, before smoothly dropping to zero for arguments greater than 1. 

To understand what the spectral action is doing, notice that functions can be applied to self-adjoint operators via their spectral decomposition
\begin{equation} f(D/\Lambda) = \sum_\lambda f(\lambda/\Lambda) |\lambda\rangle\langle\lambda| \:, \end{equation}
which in turn allows us to express the spectral action as a weighted sum over the Eigenvalues of the Dirac operator: 
\begin{align}
    S(D) = \sum_\lambda f(\lambda/\Lambda).
\end{align}
The spectral action, then, effectively counts the eigenvalues of the Dirac operator up to the cutoff $\Lambda$, with a soft weighting near the cutoff provided by the function $f$. For $f$ chosen as a smooth approximation to a step function,  the spectral action approximates the number of eigenvalues with magnitude below $\Lambda$, smoothly suppressing the contribution of higher modes. 

To make sense of how such an action might correspond to physics, it should be compared with the standard-model action coupled with the Einstein-Hilbert action,
\begin{equation} S = \int_M (\mathcal{L_{SM}}+\mathcal{L_{EH}} )\sqrt{-g}d^4x \:.  \end{equation}
This standard action computes the local energy contributions of the gravitational and matter fields at each point in spacetime, and then integrates this content over the spacetime manifold $M$ to return a number. This action provides a pointwise or `local' book keeping of the energy density across the spacetime manifold. 

The spectral action is based on the hypothesis that the physical action should only depend on the spectrum of the Dirac operator. The heuristic behind this idea is as follows: In order to calculate the action of a physical system, instead
of tracking the Lagrangian density point by point, and then integrating across spacetime to construct the physical action, one could ask what the global vibrational modes or frequencies are, and then perform a weighted sum of all to arrive at the same number. By analogy, if one wished to determine the energy in a guitar string, one could measure the curvature and tension in the string at each point, before summing all contributions across the string. Alternatively, one could determine the frequency components of the plucked string and then sum the contribution of each tone in the sound spectrum. The spectral hypothesis and the logic behind the spectral action is then based on the fact that just as a function can be analyzed in terms of its global frequency components via a Fourier transform, a Dirac operator can be analyzed via its spectral decomposition.

To make contact between the spectral action and the standard physical action, we make use of the operator-valued Laplace transform to re-express the function $f$ within the spectral action functional. It is through the Laplace transform that the action connects to heat kernel techniques, which are essential in semi-classical quantum gravity and index theory~\cite{VAN_DEN_DUNGEN_2012},
\begin{equation} f(D/\Lambda)=\int_0^\infty \tilde{f}(t) \exp(-tD^2/\Lambda^2)dt \:. \end{equation}
Note that not every function has a Laplace transform, 
but many physically reasonable functions do (this is part of the reason for $f$ being smooth, positive, and rapidly decaying).
Taking the trace, the spectral action is then expressed as an integral over heat kernels,
\begin{equation} S(D/\Lambda) =\int_0^\infty \tilde{f}(t) Tr\left[\exp(-tD^2/\Lambda^2)\right]dt \:. \end{equation}
Historically, heat kernels arose as the Green's function for the heat equation, and were used to describe the diffusion of heat. Terms of the form  $\text{Tr}\left[\exp(-tH)\right]$, where $H$ is a generalized Laplacian  have a well known expansion in $t$, known as the heat kernel expansion. In dimension 4, and where $D$ is a Dirac type operator, this allows the spectral action to be approximated for large $\Lambda$ as~\cite{VAN_DEN_DUNGEN_2012},
\begin{equation}\label{eq:heatkernel}Tr(f(D/\Lambda)) \simeq 2f_4 \Lambda^4 a_0(D^2)+ 2f_2 \Lambda^2 a_2(D^2) + f(0)  a_4(D^2) + O(\Lambda^{-2})  \:, \end{equation}
where $f_j = \int_0^\infty f(v)v^{j-1}dv$ are the moments of the function $f$ for $j>0$, and where the Seeley–DeWitt coefficients $a_k$  depend locally on geometric invariants such as curvature and gauge field strengths. Specifically:
\begin{enumerate}
    \item $a_0(D^2)$ gives the cosmological constant.
    \item     $a_2(D^2)$ gives the Einstein-Hilbert action and Higgs quadratic potential term.
    \item $a_4(D^2)$ gives Yang-Mills gauge kinetic terms, Higgs Kinetic and quartic potential terms as well as the quadratic coupling of the Higgs to Gravity and and quadratic gravity terms.
\end{enumerate}
Higher-order coefficients then correspond to higher-order correction terms in the action. For a carefully chosen spectral triple that encodes the Standard Model and its internal symmetries, the asymptotic expansion of the spectral action reproduces the bosonic part of the Standard Model Lagrangian, including gravity, with the fermionic terms $\langle \psi|D\psi\rangle$ added separately.

Strikingly, for a carefully chosen spectral triple, the spectral action yields the full bosonic Standard Model Lagrangian coupled to Einstein-Hilbert gravity~\cite{chamseddine-connes-marcolli}. The expansion also  produces higher-order curvature terms, providing testable corrections.  The framework offers a geometric origin for the Higgs field, which emerges as a component of the non-commutative connection in the internal space, and it unifies gauge and gravitational dynamics in a single geometric language, treating them on equal footing. The function $f$ also plays the role of a UV regulator, effectively smoothing out high-energy contributions in analogy with regularization in quantum field theory. Unfortunately, however, the spectral action also suffers from several conceptual and technical limitations that are the subject of ongoing research~\cite{Chamseddine_2012}. 

Perhaps the most pressing challenge with the spectral action is that it is only really well defined in Euclidean signature. More accurately, the asymptotic expansion using heat kernel techniques,  which is used to make contact with local descriptions of physics, requires ellipticity.
 In Lorentzian signature the Dirac operator is instead hyperbolic, such that standard heat kernel techniques no longer apply. The pragmatic approach is to perform all calculations in Euclidean signature before Wick rotating to recover Lorentzian physics. While this works in certain idealized situations, 
Wick rotations are not always well-defined, particularly in spacetimes with horizons, non-trivial topology, or causal obstructions to analytic continuation. Even when a Wick rotation is formally possible, it remains unclear whether the resulting Lorentzian theory captures all relevant physics. While no fully satisfactory Lorentzian formulation of the spectral action currently exists, a number of avenues are being explored, including replacements of the usual trace with structures more appropriate to Lorentzian geometry, such as Hadamard states or Krein space traces, which can accommodate operators that are not self-adjoint in the conventional sense~\cite{Nguyen,van_den_Dungen_2016,strohmaier-ncg}. Others argue that a proper treatment requires incorporating causal structure explicitly into the definition of the spectral triple itself, thus moving beyond the Euclidean setting ~\cite{Barrett_2007,Devastato_2018,bizi2017spacetimedimensionsalgebras,Franco_2014}.

Another clear limitation of the spectral action derives from arguably one of its nicest features: The action itself places additional and measurable restrictions on the physical predictions, which, although exciting, turn out not to be correct. In particular, the action is defined at a scale $\Lambda$, which is interpreted as an effective cutoff scale. On computing the heat kernel expansion of the action, one finds that an additional constraint is imposed on the gauge couplings, which is identical to the constraint that one sees at symmetry breaking in certain grand unified theories. For this reason $\Lambda$ is taken to be a unification scale~\cite{VAN_DEN_DUNGEN_2012, Chamseddine_2012}. The downside to this interpretation is that these constraints do not exactly match the observed running of couplings without further modifications~\cite{Okumura_1997}. The result is a Higgs mass prediction that lies above the observed value, necessitating refinements such as the inclusion of additional scalar fields or modified dynamics at higher scales~\cite{Chamseddine_2012}.

Finally, the action is not quantized in the traditional sense. When one performs the heat kernel expansion, one obtains a classical action, which must be quantized using all of the usual techniques. This should perhaps be no big surprise,    as the geometry that is fed into the spectral action is a classical spacetime with added internal structure. So, although the action unifies gravity and gauge interactions, the most conservative formulation that we discuss here does not yet provide a fully consistent approach to quantum gravity or renormalization. There are two approaches that one might take to producing a fully quantum theory. The first would be to quantize the model at the level of the spectral action itself, prior to performing the heat kernel expansion. This approach is explored by Rovelli, Hale, Besnard, and other researchers~\cite{hale,besquan,rov99}. Alternatively, one could replace the classical spacetime itself with a non-commutative geometry à la Seiberg and Witten~\cite{Seiberg_1999}, or other approaches~\cite{Chamseddine_2014}. However, for brevity, we will not review these approaches here.

\subsection{CFS Overview} \label{secoverview_CFS}

The CFS approach~\cite{cfs,Finster2024,website}  aims to provide a unified description of quantum theory and general relativity by reformulating physics in terms of one fundamental object:
a measure on a set of linear operators on a Hilbert space.
Instead of starting with a spacetime manifold and physical fields, spacetime and all structures therein
emerge as secondary objects from this measure.
This shift in perspective allows for a natural incorporation of both quantum and geometric aspects of physics, with the usual notions of spacetime, quantum states, and field equations arising from the underlying structure of the CFS. The framework not only unifies quantum theory and gravity at a fundamental level but also
allows for spacetimes with a non-trivial, possibly discrete microstructure on the Planck scale.

\subsubsection{CFS Overview: Fundamental Elements} \label{CFSmotivate}
Similarly to non-commutative geometry, a causal fermion system also involves a triple $(\H, \F, \rho)$. The objects of this triple have similarities with Connes' spectral triple, but there are also major differences. As in NCG, $\H$ is a Hilbert space. The set $\F$, on the other hand, does not have a direct correspondence in NCG. Given a parameter $n$ (the {\em{spin dimension}}), $\F$ is defined as the set of all self-adjoint linear operators in $\H$ of finite rank which (counting multiplicities) have at most $n$ positive and at most $n$ negative eigenvalues. In a certain sense, $\F$ is the  ambient space in which the theory is defined.
Finally, $\rho$ is 
a (positive) measure defined on a $\sigma$-algebra of subsets of $\F$. Just as with spectral triples in NCG, the elements of a CFS appear quite alien at first sight, but they are rooted in clear and elegant physical principles, with the goal of describing a spacetime together with all structures and objects therein. The benefit of formulating physics in this way is that, on the one hand, it provides novel explanations for otherwise unexplained features of the standard model including the appearance of three particle generations, while at the same time providing a natural context for quantizing gravity.

The key idea behind CFS is that the correlations between fermionic fields contain a huge amount of information not only about the causal structure of spacetime (and therefore also metric data), but also about the gauge and scalar fields in spacetime to which the fermions are coupled. The CFS formalism provides the machinery needed to reconstruct physics from this correlation data. In particular, the fermionic correlation data is packaged in the elements $(\F,\H)$, while the measure $\rho$ provides the minimal additional information required to specify the distribution of these correlations across spacetime in a way that respects both quantum mechanical principles and geometric constraints. We unpack these ideas here.

A wave function that satisfies the Dirac equation encodes causal information through its squared norm $|\psi(x)|^2= \psi(x)^\dagger \psi(x)$. 
To see this, notice that the components of a spinor field propagate at a finite speed, determined by the causal structure of spacetime. As such,  a spinor field evolved from compactly supported initial data, $|\psi(x)|^2$ vanishes outside the causal future of the initial support. In
this way, the support of $|\psi(x)|^2$ provides some information on the causal structure of
spacetime. By aggregating the information contained in all wave functions evolved from compactly supported initial
data, the complete causal structure of a spacetime can be extracted. This causal data is enough to determine the metric up to a conformal factor~\cite{hawking1976new,malament1977class}.

The self-correlations $|\psi(x)|^2$ of compactly supported solutions of the Dirac equation provide information about the causal structure and metric, but it is possible to obtain much more information~\cite{cfs} by  considering the correlations between pairs of wave functions, as described by their inner product~$\overline{\psi}(x) \phi(x)$
(where~$\overline{\psi}$ denotes the adjoint spinor; the inner product~$\overline{\psi}(x) \phi(x)$ is also
referred to as the {\em{spin inner product}} and has signature~$(2,2)$, which is indefinite). In particular, correlations between wave functions capture the relative phase and interference effects between solutions. Phase differences arise because each wave function evolves according to the Dirac equation 
\begin{align} \label{dirac}
(\slashed{D} - m)\psi = 0
\end{align}
where $\slashed{D} = i\gamma^a e_a^\mu  (\partial_\mu - \frac{i}{4}\omega_\mu + i eA_\mu)$, with spin connection $\omega_\mu$, and gauge potential $A_\mu(x)$. As wave functions propagate through spacetime in the presence of a gauge field $A_\mu(x)$, they accumulate path-dependent phase factors.
By taking into account the local information contained in \emph{all} correlations $\overline{\psi(x)}\phi(x)$, the full details of the gauge potential~$A_\mu(x)$ can be recovered up to gauge transformations (as can be seen for example by choosing~$\psi$ and~$\phi$ as
wave packets propagating in different directions; for details see~\cite[Section~5.1]{Finster2024}). A similar phase-shift effect also arises due to the Yukawa coupling between spinors and scalar fields, enabling information about the scalar sector to be extracted from the same correlations.

With the conceptual picture laid out, our next task is to explain how the correlations between physical
wave functions are captured by the data in a CFS. For simplicity, we consider a classical spacetime~$\scrM$ (being
Minkowski space or some general globally hyperbolic spacetime) to spell out how this can be encoded in a CFS.
The main idea is to describe the correlations of the wave functions at a given spacetime point~$x$
by the so-called {\em{local correlation operator}}~$F(x)$. In an orthonormal basis~$(\psi_n)$ of the Hilbert
space, the matrix entries of the local correlation operator are defined by the correlations of the
corresponding wave functions at~$x$,
\begin{equation} F(x)^i_j := -\overline{\psi_i(x)}\psi_j(x) \:. \end{equation}
In a basis-independent way, the local correlation operator can be defined by the relation
\begin{align}
    \langle{\psi}| F(x)\,\phi\rangle = -\overline{\psi(x)}\phi(x) \qquad \text{for all~$\psi, \phi \in \H$} \label{define-F}
\end{align}
(where~$\langle .|. \rangle$ denotes the scalar product on the Hilbert space~$\H$).
In short, local correlation operators are used to extract information about any two solutions at a 
spacetime point~$x$. Correlations between wave functions at two different spacetime points~$x$ and~$y$
can be obtained by considering the product of the local correlation operators~$F(x)\, F(y)$.

The local correlation operator~$F(x)$ is a symmetric operator on~$\H$ of rank at most four,
having at most two positive and at most two negative eigenvalues
(this is a consequence of the fact that the spin inner product on the right side of~\eqref{define-F}
is indefinite with signature~$(2,2)$; for details see~\cite[Chapter~5]{Finster2024}).
Thus, choosing the spin dimension~$n=2$, it is a point in the set~$\F$ of operators introduced above.
This construction gives us a distinguished set of operators~$\{ F(x)\:|\: x \in \scrM \} \subset \F$ formed of all local correlation operators. The final step in the construction of the causal fermion systems is to introduce a measure defined on this set. In our example of a classical spacetime, the manifold~$\scrM$ itself has a natural measure given by $d^4x$ (or similarly~$\sqrt{-\det g}\, d^4x$ in curved spacetime). We can use what is known as the
{\em{push-forward measure}} to transfer this measure to the space $\F$. To this end, we consider the mapping
\begin{align} \label{eq:localcorrelation}
 F&:\scrM\rightarrow \F, \qquad x \mapsto F(x),
\end{align}
which to every spacetime point associates the corresponding local correlation operator.
For a subset $A\in\F$, the push-forward measure $\rho$ is then defined by
\begin{equation} d\rho(A) =  \text{Volume of all points } x\in \scrM \text{ with } F(x) \in A. \end{equation}
The push-forward measure $\rho$ allows one to integrate over the space of correlation operators $\F$ using the original spacetime measure $d^4x$. For a function $g(F)$  defined on $\F$, the integral over $\F$ becomes:
\begin{equation} \int_\F g(F) d\rho =\int_\scrM g \big( F(x) \big) \:d^4 x \:. \end{equation}
The resulting structure~$(\H, \F, \rho)$ is the causal fermion system that describes the physical system we started with.

Before going on, we briefly comment on the significance of the measure~$\rho$.
This measure plays a double role: First, it distinguishes the local correlation operators which are
realized in our system. This is made precise by the notion of the support of the measure,
defined as the complement of the largest open set of measure zero; that is,
\begin{equation} 
\supp \rho := \F \setminus \bigcup \big\{ \text{$\Omega \subset \F$ \,\big|\,
$\Omega$ is open and~$\rho(\Omega)=0$} \big\} \:. \end{equation}
It turns out that the support is the closure of the distinguished set of local correlation operators
\begin{equation}\label{eq:spacetime} M := \supp \rho = \overline{ \{ F(x)\:|\: x \in \scrM \} } \:. \end{equation}
This makes it possible to disregard the classical spacetime~$\scrM$ and the local correlation map~$F(x)$.
Instead, we work directly with the operators in~$M$ and consider them as the points of our spacetime.
Thus, each classical spacetime point is now mapped to an operator in~$M \subset \F$ and in the context of CFS we refer to $M$ as the spacetime.
The second role of the measure~$\rho$ is that it makes it possible to associate to regions of spacetime
a corresponding volume. A measure also allows for integration. This is crucial for the formulation of
the dynamical equations of a causal fermion system in Section~\ref{seccfscyn}.

In the above example, we started with a classical four-dimensional spacetime
and constructed the corresponding CFS data $(\F,\H, \rho)$. In order to go beyond classical geometry,
one needs to go in the opposite direction: One wants to start with a CFS as a set of abstract objects,  and then to derive the corresponding spacetime, as well as all of the gauge and scalar field content. To give a flavor for how this is achieved in practice, we point out an extremely interesting, but highly non-obvious fact: First, because $F(x)$ and $F(y)$ are operators, one can consider their product $F(x)\,F(y)$, which will again be an operator
of rank at most $2n$, and with non-trivial eigenvalues $\lambda^{F(x)\,F(y)}_1,... \lambda^{F(x)\,F(y)}_{2n}$. To simplify the notation, in the following we will omit the explicit expression of the local correlation map $F(x)$ and simply label the classical spacetime point and the operator it map to by the same character, hence $x=F(x)$ when we consider a CFS that is the image of a local correlation map. In our Minkowski space example, if two points $x$ and $y$ are spacelike separated, then all the $\lambda^{xy}$ will have the same absolute value. If $x$ and $y$ are timelike separated then the $\lambda^{xy}_j$  will be all real and will not have the same absolute value, while in all other cases (i.e., if the $\lambda^{xy}_j$  are not all real and do not all have the same absolute value) then the points $x$ and $y$ are lightlike separated. The proof of this fact is rather
complicated~\cite[Section~1.2.5]{cfs}, and so we will not provide it here. 

\subsubsection{CFS Overview: Dynamics} \label{seccfscyn} \label{CFSaction}
The dynamics of a causal fermion system is described
by the {\em{causal action principle}}. It is a non-linear variational principle for the measure~$\rho$
of the causal fermion system. Before introducing the causal action principle, we collect the structures introduced above into a formal definition for a general CFS. 
\begin{Def} \label{defcfs} (causal fermion system) {\em{ 
Given a separable complex Hilbert space~$\H$ with scalar product~$\la .|. \ra_\H$
and a parameter~$n \in \N$ (the {\em{``spin dimension''}}), we let~$\F \subset \Lin(\H)$ be the set of all
symmetric operators on~$\H$ of finite rank, which (counting multiplicities) have
at most~$n$ positive and at most~$n$ negative eigenvalues. On~$\F$ we are given
a positive measure~$\rho$ (defined on a $\sigma$-algebra of subsets of~$\F$).
We refer to~$(\H, \F, \rho)$ as a {\em{causal fermion system}}.
}}
\end{Def} \noindent

In order to single out the physically admissible causal fermion systems, 
we want to minimize an action functional~$\Sact(\rho)$ under suitable variations of
the measure. More precisely, the {\em{causal action}} has the form
\begin{equation}\label{eq.Causalaction} \Sact(\rho) = \iint_{\F \times \F} \L(x,y)\: d\rho(x)\, d\rho(y) \:, \end{equation}
where~$\L$ is a non-negative function on~$\F \times \F$ defined as follows.
For any~$x, y \in \F$, the product~$x y$ is an operator of rank at most~$2n$. 
However, in general it is no longer a symmetric operator because~$(xy)^* = yx$,
and this is different from~$xy$ unless~$x$ and~$y$ commute.
As a consequence, the eigenvalues of the operator~$xy$ are in general complex.
We denote these eigenvalues counting algebraic multiplicities
by~$\lambda^{xy}_1, \ldots, \lambda^{xy}_{2n} \in \C$
(more specifically,
denoting the rank of~$xy$ by~$k \leq 2n$, we choose~$\lambda^{xy}_1, \ldots, \lambda^{xy}_{k}$ as all
the non-zero eigenvalues and set~$\lambda^{xy}_{k+1}, \ldots, \lambda^{xy}_{2n}=0$).
Then the {\em{causal Lagrangian}} is given by
\begin{equation}\label{eq:lagrangian} \L(x,y) = \frac{1}{4n} \sum_{i,j=1}^{2n} \Big( \big|\lambda^{xy}_i \big|
- \big|\lambda^{xy}_j \big| \Big)^2\:. \end{equation}
The {\em{causal action principle}} is to minimize the causal action~$\Sact$ by varying the measure~$\rho$
within the class of regular Borel measures on~$\F$ under the following constraints:
\begin{align}
\text{\em{volume constraint:}} && \rho(\F) = \text{const.} \quad\;\; & \label{volconstraint} \\
\text{\em{trace constraint:}} && \int_\F \tr(x)\: d\rho(x) = \text{const.}& \label{trconstraint} \\
\text{\em{boundedness constraint:}} && \T(\rho) := \iint_{\F \times \F} 
\bigg( \sum_{j=1}^{2n} \big|\lambda^{xy}_j \big| \bigg)^2
\: d\rho(x)\, d\rho(y) &\leq C \:, \label{Tdef}
\end{align}
where~$C$ is a given parameter (and~$\tr$ denotes the trace of a linear operator on~$\H$ of finite rank).
The constraints can be understood mathematically as being
needed in order to get a well-posed variational principle
with non-trivial minimizers.

\subsubsection{Spacetime, Causal Structure and the Physical Wave Functions} \label{CFSintrinsic}
Let~$\rho$ be a {\em{minimizing}} measure. Recall, that in \eqref{eq:spacetime} we defined {\em{Spacetime}}
as the support of this measure, $M := \supp \rho \;\subset\; \F $,  where on~$M$ we consider the topology induced by~$\F$ (generated by the operator norm
on~$\Lin(\H)$).
Thus, spacetime points are symmetric linear operators on~$\H$.
The restriction of the measure~$\rho|_M$ gives a volume measure on spacetime. We now formalize the notion of causality introduced in the section above. 

\begin{Def} (causal structure) \label{def2} 
{\em{ For any~$x, y \in \F$, we again denote the non-trivial ei\-gen\-values of the operator product~$xy$
(again counting algebraic multiplicities) by~$\lambda^{xy}_1, \ldots, \lambda^{xy}_{2n}$.
The points~$x$ and~$y$ are
called {\em{spacelike}} separated if all the~$\lambda^{xy}_j$ have the same absolute value.
They are said to be {\em{timelike}} separated if the~$\lambda^{xy}_j$ are all real and do not all
have the same absolute value.
In all other cases (i.e.\ if the~$\lambda^{xy}_j$ are not all real and do not all
have the same absolute value),
the points~$x$ and~$y$ are said to be {\em{lightlike}} separated. }}
\end{Def} \noindent
Restricting the causal structure of~$\F$ to~$M$, we get causal relations in spacetime.

The Lagrangian~\eqref{eq:lagrangian} is compatible with the above notion of causality in the
following sense.
Suppose that two points~$x, y \in M$ are spacelike separated.
Then the eigenvalues~$\lambda^{xy}_i$ all have the same absolute value.
As a consequence, the Lagrangian~\eqref{eq:lagrangian} vanishes. Thus, pairs of points with spacelike
separation do not enter the action. This can be seen in analogy to the usual notion of causality, where
points with spacelike separation cannot influence each other.
This is the reason for the notion ``causal'' in {\em{causal}} fermion system
and {\em{causal}} action principle.

In the following we will need some of the \emph{intrinsic} structures that are inherent in the definition of a CFS, e.g., for every~$x \in \F$ we define the {\em{spin space}}~$S_xM$ by~$S_xM = x(\H)$;
it is a subspace of~$\H$ of dimension at most~$2n$.
It is endowed with the {\em{spin inner product}} $\Sl .|. \Sr_x$ defined by
\begin{equation} 
\Sl u | v \Sr_x = -\la u | x v \ra_\H \qquad \text{(for all $u,v \in S_xM$)}\:. \end{equation}
A {\em{physical wave function}}~$\psi$ is defined as a function
which to every~$x \in M$ associates a vector of the corresponding spin space,
\begin{equation} \label{psirep}
\psi \::\: M \rightarrow \H \qquad \text{with} \qquad \psi(x) \in S_xM \quad \text{for all~$x \in M$}\:. \end{equation}
In this manner every vector~$u \in \H$ of the Hilbert space gives rise to a distinguished
wave function. 

\subsubsection{Analyzing the causal action principle in classical spacetimes}
In the above over-view of CFS, we explained how to construct a CFS from a classical spacetime
by encoding the spacetime structures in the correlations of Dirac wave functions
(Section~\ref{CFSmotivate}). Moreover, we 
explained how to formulate the the action principle for a CFS (Section~\ref{CFSaction}), and highlighted some of the intrinsic structures
of a CFS. 
In order to complete the picture, it remains to explain how the causal action principle can be
analyzed for causal fermion systems approximated by classical spacetimes.
In this case, the first step is to {\em{specify the fermion configuration of the vacuum}}. To this end,
in generalization of~\eqref{dirac} one formulates a Dirac equation for systems of Dirac spinors.
The way this ``system'' is built up determines which families and generations of fermionic elementary
particles (leptons and quarks) are considered. The general construction for the standard model
of elementary particles is given in~\cite[Chapter~5]{cfs}. Here we merely note that the system of
Dirac wave functions can be described by vector bundles. Algebraic structures can be introduced
or recovered by analyzing their action on the fibers of this bundle (this procedure is exemplified
for octonionic structures in~\cite{oct-cfs}).

Once the vacuum configuration has been specified, one introduces a {\em{classical
bosonic interaction}} by inserting corresponding bosonic potentials into the Dirac equation.
For example, one can couple the Dirac spinors minimally to a Yang-Mills potential or introduce
a classical gravitational field. Note that, a priori, the bosonic potentials and gravitational fields
do \emph{not} need to satisfy any field equations. This is an important point because, instead of
presuming the field equations, we want to {\em{derive}} them from the causal action principle.
To this end, one analyzes whether the causal fermion system constructed from the Dirac solutions
is a minimizer (or critical point) of the causal action principle \eqref{eq.Causalaction} or not.
On a more technical level, it is useful to work with the {\em{kernel of the fermionic projector}}~$P(x,y)$,
being a two-point correlation function constructed from the family of all physical wave functions.
Abstractly, it is given by
\begin{equation} 
P(x,y) \, \phi = -\sum_i \psi^{e_i}(x) \: \Sl \psi^{e_i}(y) |\; \phi \Sr_y \:, \end{equation}
where the~$\psi^{e_i}$ is the physical wave function (see~\eqref{psirep})
corresponding to an orthonormal basis~$(e_i)$ of the Hilbert space~$\H$.
In our concrete example of Dirac spinors, it is a bi-distributional solution of the Dirac equation.
To see its dependence on the bosonic potentials, one performs the {\em{light-cone expansion}} (in curved spacetimes usually referred to as Hadamard expansion)
\beq \begin{split}  \label{lce}
P(x,y) = & \sum_{n=-1}^\infty
\sum_{k} m^{p_k} 
{\text{(nested bounded line integrals)}} \times  T^{(n)}(x,y) \\
&+ \tilde{P}^\lec(x,y) + \tilde{P}^\hec(x,y) \:,
\end{split}
\eeq
where the nested bounded line integrals involve the bosonic potentials and their derivatives.
Here the factors~$T^{(n)}$ are singular on the light cone, and the singularity gets milder
if~$n$ is increased. Finally, $\tilde{P}^\lec(x,y)$ and~$\tilde{P}^\hec(x,y)$ are smooth contributions to the
kernel of the fermionic projector. After regularization on a small length scale $\varepsilon$
(which can be thought of as the Planck length), the EL equations of the causal action principle
can be expressed in terms of the kernel of the fermionic projector.
Analyzing these equations asymptotically for small~$\varepsilon$, one finds that
the EL equations are satisfied if and only if the potentials contained in~$\slashed{D}$ in \eqref{dirac} have a specific structure
and satisfy corresponding field equations (like, for example, Maxwell's equations).
In this way, the classical field equations can be derived from the causal action principle.
This procedure, referred to as the {\em{continuum limit}}, is worked out in detail in the textbook~\cite{cfs}.

\subsection{GTD Overview} \label{secoverview_TD}
 The GTD program aims to reformulate quantum field theory so that it is independent of classical time. The ultimate goal is a unified description of quantum gravity and particle physics that resolves the “problem of time” inherent in the canonical quantization of gravity.
GTD is based on Adler's theory of Trace Dynamics (TD) \cite{adler-trace} which is a matrix-valued Lagrangian dynamics in which the Lagrangian is the trace of a matrix polynomial. 
Owing to the existence of a novel conserved charge (the Adler–Millard charge), TD functions as a “pre-quantum” theory. Specifically, when the statistical mechanics of TD is worked out, the conserved charge enforces the canonical commutation relations, so that the averaged behavior of the matrix variables reproduces ordinary quantum mechanics. In this sense, QM is not postulated but rather emerges as the statistical equilibrium of TD.

Unlike in ordinary quantum mechanics, in TD the underlying equations of motion for the matrices are not unitary. As a result, different kinds of behavior can emerge in different regimes. When random fluctuations around equilibrium are tiny, the averaged dynamics look like standard unitary Schrödinger evolution. 
If however the fluctuations are significant, the evolution can mimic wave function collapse, recovering the sudden, stochastic jumps that are usually inserted  into quantum mechanics “by hand” to explain measurements.
 At present, however, TD is formulated only on flat Minkowski spacetime, with no inclusion of gravitation and without specification of a particular Lagrangian.

GTD removes the classical spacetime manifold from TD, replacing it by a non-commutative space labeled by the quaternions and the octonions. Gravitation,  Yang-Mills fields, and fermions are introduced into the model through the spectral action principle, and a fundamental Lagrangian is proposed
This leads to a 
    pre-spacetime, pre-quantum candidate theory of unification, from which general relativity, quantum field theory, and the standard model of particle physics emerge, with one possible gauge group of the unified theory being $\text{E}_8 \otimes \text{E}_8$.

\subsubsection{TD Overview: Fundamental Elements} 

Among the foundational problems of quantum theory, perhaps the most significant is the problem of time. Our universe today is predominantly classical and consists mostly of macroscopic objects such as stars and galaxies. It is for this reason that the universe can be described by a classical spacetime manifold, which in turn provides the classical time parameter used to describe evolution in quantum theory. If the universe were not dominated by classical objects, there would be no classical spacetime manifold and no time parameter to use in quantum theory. This is true even at the currently accessible low energies, and is a direct consequence of the Einstein hole argument~\cite{Carlip}.

It follows that the current formulation of quantum field theory, defined relative to classical spacetime coordinates, is contingent on the universe being dominated by classical bodies. Such bodies are a limiting case of quantum systems, and hence quantum theory depends on its own limit. Even at low energies, this can only be an approximate formulation. The low-energy universe could in principle be entirely devoid of macroscopic objects and hence devoid of classical time. In other words, the fundamental building blocks admit no continuum approximation. However, one should still be able to describe the dynamics of microscopic systems. Hence, there must exist a reformulation of quantum field theory which does not depend on classical time

Generalized Trace Dynamics (GTD) is a pre-spacetime, pre-quantum theory that also provides a framework for quantum gravity, in the sense that gravitation is matrix-valued and non-classical, with classical gravity emerging as an approximation. GTD further aims to address several foundational problems in quantum theory:
\begin{itemize}
\item Why does the wave function collapse during a quantum measurement?
\item Why are the outcomes of a quantum measurement random, even though the Schrödinger equation is deterministic? Why do these random outcomes obey the Born probability rule?
\item How do entangled quantum systems maintain correlations outside the light cone? Does this indicate an incomplete understanding of spacetime structure and special relativity?
\item Why does relativity allow supra-quantum nonlocal correlations that exceed the Tsirelson bound~\cite{Rabsan}?
\item The standard model contains over twenty dimensionless fundamental constants, whose numerical values are unexplained. GTD aims to derive these constants from first principles, addressing existing low-energy data without requiring higher-energy experiments.
\item GTD should make testable experimental predictions, ensuring the theory is falsifiable.
\end{itemize}

We progress towards developing GTD by building on Adler's theory of trace dynamics, which we summarize next.

\subsubsection{Trace dynamics}\label{sec:TD}

Conventional quantization in the Heisenberg picture proceeds as follows. Given a Hamiltonian dynamics, one raises classical dynamical variables to the status of matrices/operators and replaces the Poisson brackets by Heisenberg commutation relations. Hamilton’s equations of motion are replaced by Heisenberg's equations of motion.

In contrast, trace dynamics seeks to derive quantum theory from an underlying more general dynamics, rather than  by `quantizing' a classical dynamics wherein the quantum commutation relations are imposed in an ad hoc manner. In TD one raises classical dynamical variables to matrices but does not impose Heisenberg's commutation relations on these matrices. Instead, one constructs a Lagrangian / Hamiltonian dynamics with these matrix-valued dynamical variables.

For illustration, consider the Lagrangian for a collection of non-relativistic free particles
\begin{equation} S = \sum_i\;  \int d\tau \; \frac{1}{2}m_i\left(\frac{dq_i}{d\tau}\right)^2 \:. \end{equation}

To proceed to trace dynamics, make the transition from real numbered dynamical variables to matrices: 
$q\rightarrow {\bf q}$ and replace the Lagrangian by a trace of the matrix polynomial $\mathcal{L}$,
\begin{equation} S = \sum_i\;  \int d\tau \; \text{Tr}\left[ \frac{1}{2}\frac{L_{p}^2}{L^2}\left(\frac{d{\bf q}_i}{d\tau}\right)^2 \right] \:. \end{equation}
Here, we have replaced the masses by a length scale $L$, in view of subsequent applications.
 In order to vary the trace Lagrangian with respect to an operator, the notion of a trace derivative is introduced. The derivative of the trace Lagrangian ${\bf L}\equiv \text{Tr}{\mathcal L}$  with respect to an operator $\mathcal{O}$ in the polynomial ${\mathcal{L}}$ is defined as
\begin{equation} \delta {\bf L }= \text{Tr} \frac{\delta{\bf L}}{\delta\mathcal{O}}\delta\mathcal{O} \:. \end{equation}
This so-called trace derivative is obtained by varying ${\bf L}$ with respect to ${\mathcal O}$ and then cyclically permuting ${\mathcal O}$ within the trace, so that $\delta\mathcal O$ sits to the right of the polynomial $\mathcal{L}$. 
 The action can now be varied with respect to the matrix-valued variables to arrive at the familiar equations of motion and the familiar definition of canonical momentum
\begin{equation} \ddot{\bf q}_i =0\; , \qquad {\bf p}_i=\frac{L_P^2}{L^2}\; \dot{\bf q}_i \:. \end{equation}
An equivalent Hamiltonian dynamics as well as Poisson brackets can be constructed in the standard way. The matrix-valued dynamical variables do not commute with each other; in fact, the commutator of any two variables evolves dynamically, as one would expect. The commutator $[{\bf q}, {\bf p}]$ is not frozen at the Heisenberg value $i\hbar$. Evolution in TD is, in general, non-unitary, unlike the Heisenberg evolution in quantum theory.

In TD, the entries in the matrices are Grassmann numbers~\cite{adler-trace} over the field of complex numbers, keeping in mind that one has to make contact with quantum field theory. Grassmann numbers can be even(odd) grade, depending on whether they are a product of an even(odd) number of Grassmann elements. Every matrix can be written as a sum of two matrices, one consisting only of even-grade Grassmann numbers as entries, and the other consisting only of odd-grade Grassmann numbers. The former are known as bosonic matrices and are denoted by a subscript $B$ (e.g. $({\bf q}_B, {\bf p}_B))$, while the latter are called fermionic matrices and are denoted by the subscript $F$ (e.g. $({\bf q}_F, {\bf p}_F))$. By definition, bosonic(fermionic) matrices describe bosons(fermions). The fact that bosons(fermions) have integral(half-integral) spin must be derived from first principles, in trace dynamics.

One does not quantize trace dynamics. It is already a pre-quantum theory. This is because the Lagrangian is a trace over a matrix polynomial. Because the trace is invariant under cyclic permutations, the trace Lagrangian as well as the trace Hamiltonian are invariant under a global unitary transformation of the dynamical variables. This gives rise to a novel conserved charge, known as the Adler-Millard charge~\cite{adler-trace}, which is conserved under the internal evolution of the matrix variables in trace dynamics.
\begin{equation}\label{eq:adler-miller-charge}
    \tilde{C} = \sum_{r\in B} \; [{\bf q}_{Br}, {\bf p}_{Br}] -\sum_{r\in F}\left\{{\bf q}_{Fr},{\bf p}_{Fr}\right\} \, .
\end{equation}
An important assumption in deriving this charge is that any constant coefficients in the Lagrangian must be $c-$numbers, not matrix-valued.

This charge, denoted as $\tilde{C}$ and named after its discoverers, has dimensions of action and is absent in classical dynamics. This charge is what makes TD different from both classical dynamics and quantum dynamics. Quantum dynamics is in between classical dynamics and TD, being more general than the former, and less general than the latter. The commutator of observables in quantum mechanics has the physical dimensions of an action. We see that even though the values of individual commutators evolve dynamically, their sum is conserved during evolution. Therefore, the commutators of the various degrees of freedom are exchanging a quantity of dimension action among each other when the system is evolving dynamically. Quantum field theory is that emergent approximation of TD in which such an exchange no longer takes place (equipartition), and each commutator gets frozen at the Heisenberg value. See~\cite[Chapter 5]{adler-trace} for the emergence of quantum field dynamics as a result of equipartition. 

Trace Dynamics  is a matrix-valued Hamiltonian/Lagrangian dynamics that is Poin\-caré invariant (when formulated on flat Minkowski spacetime). It is deterministic, generally non-unitary, but norm-preserving. For a discussion of Poincar\'e invariance, see  \cite[Section 2.3]{adler-trace}, and for non-unitarity of TD, see \cite[Section 1.5]{adler-trace}. 
 Norm preservation is an explicit assumption in TD,
 which can be justified in generalized trace dynamics (GTD). 

Assuming that TD describes physics at some fundamental time resolution  $\tau_P$,  below scales tested in the laboratory, a natural question is the form of the emergent dynamics at coarser time resolution  $\tau\gg \tau_P$, accessible to experiments. This question is answered by applying the standard techniques of statistical thermodynamics, i.e. by coarse-graining the trajectories in phase space to a time-scale $\tau$ and constructing a probability distribution function. The emergent dynamics at equilibrium is determined by maximizing the Boltzmann entropy over the averaged microscopic states. There are two possibilities for the emergent dynamics. 

\textit{Case 1:} The Hamiltonian of the theory can be well-approximated to be self-adjoint, and the Adler-Millard charge is consequently anti-self-adjoint. In this case this novel charge gets equipartitioned and the Heisenberg commutation relations of quantum theory emerge for canonical averages of dynamical variables:
\begin{equation} [q_B, p_B]=i\hbar \ \ ; \quad \{q_F, p_F\} = i\hbar \:. \end{equation}
In this same approximation, the Heisenberg equations of motion, and the rules of quantum field theory,
emerge at statistical thermodynamic equilibrium. The reader is referred to Section 5 of Adler's book \cite{adler-trace} for further details. In this sense, quantum field theory is said to be an emergent phenomenon.

\textit{Case 2:} The anti-self-adjoint part of the Hamiltonian is significant, and the evolution is dominantly non-unitary. Now, the Adler-Millard charge cannot be taken to be anti-self-adjoint. Non-unitary evolution in combination with norm preservation leads to a non-linear Schrödinger equation, and the
evolution then breaks linear quantum superposition. Coarse-graining the dynamics introduces an apparent randomness in the evolution, while obeying Born's
rule: this is the phenomenon of spontaneous localization. In this way, some progress towards understanding the origin of the quantum-to-classical transition is made, though more research is needed to arrive at a precisely formulated theory. 

In summary, the underlying theory of trace dynamics is deterministic and non-unitary. If non-unitarity is significant, the linear superposition is broken. Upon coarse-graining, the emergent theory is either quantum theory or classical dynamics, depending on whether or not the Hamiltonian is dominantly self-adjoint. We conclude: Quantum theory is an emergent approximation to an underlying deterministic theory which is non-unitary but norm-preserving. 
Randomness is due to coarse graining over the time of arrival \cite{Kakade_2023}.

We also note that trace dynamics admits supra-quantum nonlocal correlations and violates the Tsirelson bound in the CHSH inequality~\cite{Rabsan}. This in itself suggests that quantum theory is approximate, not exact, and that the aforesaid violation of the Tsirelson bound should be looked for in Bell-type experiments.

Trace dynamics does not specify any fundamental Lagrangian, and while it is  a pre-quantum theory, it is not a pre-spacetime theory. Spacetime is assumed to be classical and flat (no gravitation). But this is an intermediate step towards developing a pre-spacetime, pre-quantum theory. We now propose to include matrix-valued gravitation, and a non-commutative pre-spacetime and also propose a fundamental Lagrangian \cite{Singh_2023}. 
By matrix-valued gravitation we mean, in the sense of TD, that the dynamical variables of gravity are now matrices. By non-commutative pre-spacetime we mean that every (real-number valued) spacetime point of the classical spacetime manifold is replaced by a quaternion/octonion.
This leads to a candidate theory for quantum gravity, and for unification of fundamental forces. Thus, generalized trace dynamics is where we achieve the desired result of a quantum theory without classical time, by replacing the pseudo-Riemannian spacetime manifold by the non-commutative space of quaternions / octonions, and replacing classical gravitation by a matrix-valued gravitation. The dynamical rules continue to be those of trace dynamics.

\subsubsection{Generalized trace dynamics}
The transition from classical dynamics to generalized trace dynamics (GTD) is quite a natural one. Classical dynamics builds on spacetime labeled by real numbers, on which there live dynamical variables, also labeled by real numbers or in mathematical terms, spacetime is a manifold that is locally homeomorphic to $\R^n$ and the metric has coefficients in the real numbers. Matter / field degrees of freedom locally obey the external symmetry of Lorentz invariance, and the internal Yang-Mills gauge symmetries of the standard model.  To move from classical dynamics to GTD, the  spacetime points are enriched. Instead of being just labeled by real numbers, more complex number systems enlarge both the representation of matter fields and the structure of spacetime itself. 

The quaternions, together with spinors from the Clifford algebra $\text{Cl}(2)$, suffice to describe chiral leptons of a single handedness.  Split bi-quaternions and the associated $\text{Cl}(3)$ extend this to leptons of both chiralities. The Octonions, with $\text{Cl}(6)$, can be used to include quarks: they describe one generation of quarks and leptons of one handedness. Split bi-octonions and $\text{Cl}(7)$ then account for both chiralities, while to capture three generations, one uses $\text{Cl}(8)$ and the exceptional Jordan algebra $J_3(O_C)$ for a single chirality, and $\text{Cl}(8)\oplus \text{Cl}(8)$ with two copies of $J_3(O_C)$ for both \cite{Vaibhav_2023, singh2025fermionmassratiosexceptional}.

On the spacetime side, these same algebras reshape geometry: a quaternion corresponds to 4-dimensional spacetime, a bi-quaternion to two coupled copies of 4D spacetime embedded in six dimensions, and octonions extend spacetime so as to incorporate internal symmetry spaces alongside the external dimensions.
A key question is then what  the appropriate classical gravitational variables are that should be raised to the status of matrices in TD, and which will hence describe quantum gravity? It turns out, for various reasons, that the components of the spacetime metric are not the appropriate choice. Instead, the appropriate choice is indicated by the celebrated spectral action principle of Chamseddine and Connes \cite{Chamseddine_1997}; see Section \ref{sec:spectralaction} above. According to this principle, the Einstein-Hilbert action of general relativity can be expressed in terms of the eigenvalues $\lambda_i$ of the Dirac operator $D$ on the Riemannian manifold, using the so-called truncated heat kernel expansion (this corresponds to  Eq.~(\ref{eq:heatkernel}) with $\Lambda\propto L_P^{-1}$)

\begin{equation} \text{Tr} [L_P^2 D^2] \sim L_P^{-2}\int d^4x\; \sqrt{g}\; R + {\mathcal O}(L_P^{0}) =\sum_{i}\lambda_i^2  \:. \end{equation}
These eigenvalues serve the role of dynamical observables of general relativity, as an alternative to the metric, as has been shown by Landi and Rovelli~\cite{Landi_1997}. For the purpose of incorporating gravitation into trace dynamics, every such eigenvalue $\lambda_i$ will be raised to the status of a matrix, to be denoted $\dot{q}_{Bi}$. The dot here indicates a time derivative, not with respect to the coordinate time of the classical spacetime manifold, but with respect to the so-called Connes time $\tau$. The point is that when we replace the points of the spacetime manifold by quaternions, and along-with raise Dirac eigenvalues to the status of matrices, we are in the domain of NCG. Connes' NCG admits, by virtue of the Tomita-Takesaki theorem \cite{connes} for von Neumann algebras, a one-parameter family of outer automorphisms, which play the role of a time parameter. This feature is unique to NCG and absent in Riemannian geometry. In this regard, there is a striking parallel between Connes time and the Adler-Millard charge $\tilde C$: both are a consequence of non-commutativity of the algebra, and likely bear a subtle relationship to each other.
It needs to be explored if in TD, phase space evolution taking place along the path which preserves $\tilde C$ defines a time parameter and if so, is that parameter identical to Connes time?

The spectral action principle can be generalized to include Yang-Mills fields coupled to gravity. Now, the trace of the square of the generalized Dirac operator relates to the action for gravity plus the action for gauge fields. We will raise these generalized Dirac eigenvalues to matrices in TD, denoting them by $\dot{q}_{Bi} + q_{Bi}$, where we have identified the YM dynamical variable with $q_B$. Going further, the Dirac action for fermions coupled to gravity and YM can also be expressed in terms of eigenvalues of
$D$ - these fermionic degrees of freedom are also raised to the status of matrices in TD. See Section 2.3.4 below for details. In order to make contact with the standard model and general relativity, one must specify the gauge symmetries of the theory.
In the $\text{E}_8 \otimes \text{E}_8$ unification theory \cite{kaushik_2024}, the existence of three generations of standard model quarks and leptons is proved from first principles. The fermionic matrices for leptons will be denoted $\dot{q}_{Fi}$ and those for quarks will be denoted $q_{Fi}$. A proposed branching and symmetry breaking lead to the emergence of standard-model forces and of general relativity, alongside two newly predicted interactions.

The operator $\dot{q}_{Bi}$ is identified with the very same Dirac operator $D$ whose eigenvalue $\lambda_i$ was raised to $\dot{q}_{Bi}$. We explain this with some care by bringing in the quaternions and split bi-quaternions. The quaternions give rise to the Clifford algebra $\text{Cl}(0,3)$, which has eight vectors. The algebra can be represented as
$(\mathbb{H}+\omega\mathbb{H})$ with \(\omega=e_1e_2e_3\) playing the role of the split imaginary satisfying \(\tilde{\omega}=-\omega\), \(\omega^2=1\). The imaginary quaternions associated with each set of quaternions are \((\hat{i}, \hat{j}, \hat{k}\)) and (\(\hat{l}, \hat{m}, \hat{n}\)). The elements inside each set anti-commute with each other, whereas the elements of one set commute with the elements of the other set. These can be used separately as vectors in 3D spaces, and here we use them to describe a 6D spacetime as follows
\begin{equation}
x_6=t_1\hat{i}+t_2\hat{j}+t_3\hat{k}+\omega(x_1\hat{l}+x_2\hat{m}+x_3\hat{n}) \, .
\end{equation}This gives the square modulus
\begin{equation}
x_6\tilde{x_6}=t_1^2+t_2^2+t_3^2-x_1^2-x_2^2-x_3^2
\end{equation} for the Minkowski spacetime SO(3,3). 
We construct the Dirac operator 
\begin{equation}
D_6=\hat{i}\partial_{t1}+\hat{j}\partial_{t2}+\hat{k}\partial_{t3}+\omega(\hat{l}\partial_{x1}+\hat{m}\partial_{x2}+\hat{n}\partial_{x3}) \equiv D_L + \omega D_R \, . 
\end{equation} Hence, 
\begin{equation}
D_6\tilde{D_6}=\partial_{t1}^2+\partial_{t2}^2+\partial_{t3}^2-\partial_{x1}^2-\partial_{x2}^2-\partial_{x3}^2
\end{equation}with the 6D Dirac equation being
\begin{equation}ihD_6\psi=Qc\psi \, .
\end{equation}This equation reduces to the standard Dirac equation in 6D after mapping these split bi-quaternions to gamma matrices, as follows from the work of Wilson \cite{Wilson_2022}. 

A 6D spacetime with signature $(3, 3)$ contains as subspaces a pair of 4D Minkowski spacetimes with relatively flipped signatures $(1,3)$ and $(3,1)$. Consequently, $D_6$  can be decomposed into two 4D Dirac operators \(D_4\) and \(D_4'\)
\begin{equation} D_4=\hat{i}\partial_{t1}+\omega(\hat{l}\partial_{x1}+\hat{m}\partial_{x2}+\hat{n}\partial_{x3}) \:, \end{equation}
which gives 
\begin{equation}
D_4\tilde{D_4}=\partial_{t1}^2-\partial_{x1}^2-\partial_{x2}^2-\partial_{x3}^2 \, .
\end{equation}Similarly for the Dirac operator \(D_4'\), we can write 
\begin{equation}
D_4'=\omega \hat{l}\partial_{x1}+\hat{i}\partial_{t1}+\hat{j}\partial_{t2}+\hat{k}\partial_{t3} \, . \end{equation}This gives the correct signature.
Both these Dirac operators relate to the gamma matrices using Wilson's work since \(SO(3,3) \sim \SL(4,\mathbb{R})\) and the set of generators of \(SL(4,\mathbb{R})\) has two subsets for two 4D spacetimes. Wilson \cite{Wilson_2022} showed a map from these vectors and bi-vectors to gamma matrices. We can also use this to reduce this quaternionic Dirac operator to the usual one. We employ \(\hat{i} \mapsto \gamma^0,\omega\hat{l} \mapsto\gamma^1,\omega\hat{m}\mapsto\gamma^2,\omega\hat{n}\mapsto\gamma^3\) to get the usual Dirac equation.

Returning to TD, we identify the variable $\dot{q}_{BR}$ with the curved version of the Dirac operator $D_R$, and the variable $\dot{q}_{BL}$ with the curved version of the Dirac operator $D_L$. 
That is, we define ,
\begin{equation*}
    \dot{q}_{BR} - \dot{q}_{Bt1}\hat{i}\partial_{t1}= D_{Rcurved}=
\omega(\dot{q}_{BRx1}\hat{l}\partial_{x1}+\dot{q}_{BRx2}\hat{m}\partial_{x2}+\dot{q}_{BRx3}\hat{n}\partial_{x3})
\end{equation*}
and we define an analogous expression for $\dot{q}_{BL}$. Then, bringing the left-handed and right-handed part together, we define $\dot {q}_{BLR} \equiv \dot{q}_{BL} + \dot{q}_{BR}$.
In this manner, the matrix-valued components of $\dot{q}_{BR}$ capture information about gravitation and are TD analogs of the metric. The associated gauge symmetry is $\text{SU}(2)_R\otimes U(1)_{YDEM}$. 

The components of $\dot{q}_{BL}$ capture information about the electroweak force, which is to be viewed as an external spacetime symmetry associated with the second 4D spacetime, and has the equivalent gauge symmetry $\text{SU}(2)_L \otimes U(1)_Y$. Prior to a chiral symmetry breaking (identified with electroweak symmetry breaking), the two 4D spacetimes unify into the aforementioned 6D spacetime, and this is accompanied with gravi-weak unification into $\text{SU}(3)\otimes \text{SU}(3)$ symmetry. 

We define the variables $\dot{q}_{FL}$ to describe left-handed leptons, and $\dot{q}_{FR}$ to describe right-handed leptons, and together $\dot{q}_{FLR}\equiv\dot{q}_{FL} + \dot{q}_{FR} $. The spinor states of one generation of chiral leptons are constructed from the complexified Clifford algebra $\text{Cl}(3)$ \cite{Vaibhav_2023}. 

The quantum-to-classical transition and the recovery of classical spacetime is enabled by spontaneous localization of entangled degrees of freedom. In this process, a matrix is mapped back to one of its eigenvalues - this is the inverse of the former process of TD quantization where eigenvalues are replaced by matrices. For a detailed explanation, see Section IV of \cite{maithresh_2020}.

 In summary, the Clifford algebra $\text{Cl}(3)$ obtained from the complex quaternions, and the associated symmetries $\text{SU}(2)_L\otimes U(1)_Y$ and $\text{SU}(2)_R\otimes U(1)_{YDEM}$ are adequate for describing electroweak interactions, gravitation, and one generation of chiral leptons. In the language of fiber bundles, we have a tangent bundle (a vector bundle) for a (3+3) Lorentzian manifold broken into two overlapping 4D spacetimes with relatively flipped signatures. After symmetry breaking these two 4D spacetimes have their respective tangent bundles, and respective associated vector bundles for chiral leptons.

To bring in quarks, the strong force, and three fermion generations, we must appeal to complex split bi-octonions and the associated Clifford algebra $\text{Cl}(7)$. The additional four directions provided by the octonion add on to the quaternion, and provide a principal bundle structure on top of the tangent bundle of the flipped 4D spacetime. The associated 4D Dirac operator generalizes from $D_4' $ to $D_4'  + q_{BL}$. The newly introduced $q_{BL}$ variables describe QCD, and the $\text{E}_8 \otimes \text{E}_8$ theory shows that the associated gauge symmetry is $\text{SU}(3)_c$. Correspondingly, $D_4$ is generalized to $D_4  + q_{BR}$. The newly introduced variables $q_{BR}$ describe a new force $\text{SU}(3)_{grav}$, and $q_B \equiv q_{BL} + q_{BR}$.
The variables $q_{FL}$ describe left-handed quarks, while $q_{FR}$ describes right-handed quarks, and $q_{FLR} = q_{FL} + q_{FR}$. The Clifford algebra $\text{Cl}(7)\sim \text{Cl}(6) \oplus \text{Cl}(6)$ associated with complexified split bi-octonions describes one generation of chiral quarks and leptons. An $\text{SU}(3)_{genL}\otimes \text{SU}(3)_{genR}$ symmetry arising from the decomposition of $\text{E}_8 \otimes \text{E}_8$ enables the existence of three generations of chiral fermions \cite{kaushik_2024, singh2025fermionmassratiosexceptional}. Finally, the breaking of an $\text{SU}(3)_{ST1}\otimes \omega \text{SU}(3)_{ST2}$ symmetry gives rise to the fiber bundle for the 6D fiber space (4D + 4D quaternionic) overlaid by the principal bundle for strong forces (octonionic) \cite{SinghST2025}. The quaternions are enough for spacetime and for gravi-weak, whereas octonions are needed for QCD and $\text{SU}(3)_{grav}$ with these last two symmetries being internal and connected to the principal bundle on the spacetime. The chiral symmetry breaking separates the tangent space (spacetime) from the fiber of the principal bundle (internal gauge symmetries). The internal symmetries are those which are unbroken.

The act of raising Dirac eigenvalues to Dirac operators in TD is akin to quantization. Crucially, this is accompanied by raising each spacetime point to a quaternion (for spacetime) and then to an octonion (for including internal symmetries). This latter step is absent in quantum field theory but is necessary, at all energy scales, if we are to obtain a reformulated quantum field theory, which does not refer to classical time. Spacetime in GTD is an operator-valued space, being a collection of quaternions/octonions, instead of being a collection of real numbers.

The opposite of quantization is spontaneous localization, where the Dirac operators are sent to distinct eigenvalues, and every quaternion is sent to a distinct real number. The collection of these real numbers constitutes a Lorentz-invariant tangent bundle for a Minkowski spacetime, and the spectral action principle applied to the emergent eigenvalues gives rise to the action principle of classical physics \cite{Singh_2023}.

To put things into perspective, we recall that the fundamental symmetry $\text{E}_8 \otimes \text{E}_8$ is broken as follows. Each of the $\text{E}_8$ decomposes into four $\text{SU}(3)$s, and the interpretations of these eight $\text{SU}(3)$s are the following. For the first $\text{E}_8$
\begin{equation} \text{E}_8 \rightarrow \text{SU}(3)_{ST1}\otimes \text{E}_6 \rightarrow \text{SU}(3)_{ST1}\times \text{SU}(3)_{genL} \otimes \text{SU}(3)_c \otimes \text{SU}(2)_L \otimes \text{U}(1)_Y \:. \end{equation}
For the second $\text{E}_8$
\begin{equation} \text{E}_8 \rightarrow \text{SU}(3)_{ST2}\otimes \text{E}_6 \rightarrow \text{SU}(3)_{ST2}  \times  \text{SU}(3)_{genR} \otimes \text{SU}(3)_{grav} \otimes \text{SU}(2)_R \otimes \text{U}(1)_{YDEM} \:. \end{equation}
The three forces of the standard model are recovered, along with general relativity and two newly predicted forces. Three generations of standard-model fermions are recovered, along with three types of right-handed sterile neutrinos. There are two Higgs bosons in the theory: the SM Higgs and a newly predicted one; both the Higgs are composite.

In Kaluza-Klein theories, it is known that two additional dimensions are required to unify gravity with the weak force, and an additional four are needed for unifying it with the strong force. It is highly encouraging that, in our approach to unification, these additional dimensions do not have to be imposed in an ad hoc manner but arise naturally from the dimensions of quaternions and octonions.

Of the eight $\text{SU}(3)$s, four are associated with internal symmetries, these being $\text{SU}(3)_{genL} \otimes \text{SU}(3)_c$ and $\text{SU}(3)_{genR}\otimes \text{SU}(3)_{grav}$. The other four are associated with external symmetries, these being $\text{SU}(3)_{sT1}\otimes \omega \text{SU}(3)_{ST2}$ and $\text{SU}(3)\otimes \text{SU}(3)\rightarrow \text{SU}(2)_L\otimes \text{U}(1)_Y \otimes \text{SU}(2)_R\otimes \text{U}(1)_{YDEM}$. The spacetime pair of $\text{SU}(3)$s also undergoes symmetry breaking, giving rise to 6D spacetime and to the internal symmetry space \cite{SinghST2025}.

We can now proceed to write down the action for the theory, which will be defined on the sixteen-dimensional split bi-octonionic space. We define the dynamical variable $\dot{q}_B$ on 8D split octonionic space as
\begin{equation} \dot{q}_B = \dot{q}_{B0L} + \omega \dot{q}_{B0R} + \dot{q}_{BL} + \omega\dot{q}_{BR} \:. \end{equation}
These are eight components in total. Next, we define the dynamical variable $q_B$ on 8D split octonionic space as follows,
\begin{equation} q_B = q_{BL} + \omega q_{BR} \:. \end{equation}

We define the fermionic variables $\dot{q}_{F}$ and $q_F$, each on 8D split octonionic space by
\begin{equation} \dot{q}_F = \dot{q}_{F0L} + \omega \dot{q}_{F0R} + \dot{q}_{FL} + \omega\dot{q}_{FR} \end{equation}
and
\begin{equation} q_F = q_{FL} + \omega q_{FR} \:. \end{equation}
The matrix variables $\dot{q}_{Bi}$ are atoms of spacetime - they are the microscopic degrees of freedom whose entanglement followed by spontaneous localization gives rise to classical spacetime. We also know that any matrix $q_i$ can be written as a sum of a bosonic matrix and a fermionic matrix: $q_i=q_{Bi} + q_{Fi}$ and $\dot{q}_i = \dot{q}_{Bi} + \dot{q}_{Fi}$. We call such matrices $q_i$  ``atoms of spacetime-matter'' (STM). The action that we will construct for the theory will be an action for such atoms. An STM atom is a fermion along with all the bosonic fields to which it gives rise. Thus, an electron along with its electromagnetic, weak, and gravitational fields is an example of an STM atom.

\subsubsection{GTD Overview: Dynamics} 
We illustrate the construction of the action in successive steps. Let us begin with general relativity on 4D spacetime. Following the form of the Dirac operator $D_4$ written in the previous section, we can define the matrix variable
\begin{equation} \dot{q}_{BR} \equiv D_{4curved} = \dot{q}_{Bt1}\hat{i}\partial_{t1}+\omega(\dot{q}_{Bx1}\hat{l}\partial_{x1}+\dot{q}_{Bx2}\hat{m}\partial_{x2}+\dot{q}_{Bx3}\hat{n}\partial_{x3}) \:. \end{equation}
The trace dynamics action for this gravitational matrix variable (a single atom of space-time) is
\begin{equation} S = \int d\tau \; \text{Tr}\; [\dot {q}_{BR}^2] \:, \end{equation}
and it is motivated by the spectral action principle. By summing this over all atoms of spacetime, we get the pre-spacetime, pre-quantum GTD action for quantum gravity
\begin{equation} S = \sum_i \int d\tau \; \text{Tr}\; [\dot {q}_{BRi}^2] \:. \end{equation}
Spontaneous localization will send the $i$-th atom to the $i$-th eigenvalue of the Dirac operator. Correspondingly, the underlying quaternion associated with pre-spacetime is localized to a real-number-valued point, and the collection of points (coming from localization of a collection of quaternions)  defines the emergent classical spacetime. The collection of eigenvalues gives rise to the Einstein-Hilbert action for general relativity, via the spectral action principle \cite{maithresh_2020}.

We can incorporate right-handed leptons $\dot{q}_{FL}$ into this gravity action and get an action principle for right-handed STM atoms, now written in dimensionless form,
\begin{equation} \frac{S}{\hbar}= \int \frac{d\tau}{\tau_{Pl}}\; \text{Tr} \bigg\{ \frac{L_P^2}{L^2}\left[\dot{q}_{BR}^\dagger +  \frac{L_P^2}{L^2}\beta_1 \dot{q}_{FR}^\dagger \right ] \times \left[\dot{q}_{BR} + \frac{L_P^2}{L^2} \beta_2 \dot{q}_{FR}\right] \bigg\} \:. \end{equation}
Here, $\beta_1$ and $\beta_2$ are two odd-grade Grassmann elements, included to make the trace Lagrangian bosonic. They must be unequal, $\beta_1 \neq \beta_2$, to achieve consistent equations of motion. The associated gauge symmetry is $\text{SU}(2)_R \otimes \text{U}(1)_{YDEM}$.

An analogous GTD action principle can be set up for the electroweak interaction, using the Dirac operator $D_4'$ on the second 4D spacetime. The right-handed $R$-valued variables are replaced by left-handed $L$ valued variables, to obtain the action for the electroweak interaction coupled to left-handed leptons
\begin{equation} \frac{S}{\hbar}= \int \frac{d\tau}{\tau_{Pl}}\; \text{Tr} \bigg\{ \frac{L_P^2}{L^2}\left[\dot{q}_{BL}^\dagger +  \frac{L_P^2}{L^2}\beta_1 \dot{q}_{FL}^\dagger \right ] \times \left[\dot{q}_{BL} + \frac{L_P^2}{L^2} \beta_2 \dot{q}_{FL}\right] \bigg\} \:. \end{equation}
The associated gauge symmetry is $\text{SU}(2)_L \otimes \text{U}(1)_Y$. The idea here is that the electroweak sector is the
left-handed counterpart of the gravi-DEM
sector $SU(2)_R \times U(1)_{YDEM}$ (where DEM stands for
``dark electromagnetism'').

Gravi-weak unification consists of combining the left-handed and right-handed variables and writing the GTD action on split bi-quaternionic space and the associated 6D spacetime with signature $(3,3)$,
\begin{equation} \frac{S}{\hbar}= \int \frac{d\tau}{\tau_{Pl}}\; \text{Tr} \bigg\{ \frac{L_P^2}{L^2}\left[\dot{q}_B^\dagger +  \frac{L_P^2}{L^2}\beta_1 \dot{q}_F^\dagger \right ] \times \left[\dot{q}_B + \frac{L_P^2}{L^2} \beta_2 \dot{q}_F\right] \bigg\} \:. \end{equation}
 The associated gauge symmetry is $\text{SU}(3)\otimes \text{SU}(3)$. Its spontaneous symmetry breaking gives rise to general relativity and to the electroweak interaction on the two respective 4D spacetimes. Understanding this symmetry breaking is research currently in progress \cite{graviweak2025}.

 By incorporating undotted variables and going over to the 16D split bi-octonionic space, we can bring in (left-handed and right-handed) quarks and the two new (unbroken) symmetries $\text{SU}(3)_c$ and $\text{SU}(3)_{grav}$. These two are internal symmetries. Hence,
 \begin{eqnarray*}
     \frac{S}{\hbar}= \int \frac{d\tau}{\tau_{Pl}}\; \text{Tr} \bigg\{ \frac{L_P^2}{L^2}\left[\left(\dot{q}_B^\dagger + i\frac{\alpha}{L} q_B^\dagger\right) + \frac{L_P^2}{L^2}\beta_1\left( \dot{q}_F^\dagger + i\frac{\alpha}{L}q_F^\dagger\right)\right ] \times \\ \left[\left(\dot{q}_B + i\frac{\alpha}{L}q_B\right) + \frac{L_P^2}{L^2} \beta_2 \left(\dot{q}_F+ i\frac{\alpha}{L}q_F\right)\right] \bigg\} \, .
 \end{eqnarray*}
 This is the fundamental form of the action of the theory after the $\text{E}_8\times \text{E}_8$ symmetry is broken into eight $\text{SU}(3)$s. 

 The equations of motion can be read out after first defining
\begin{equation} q_1 = q_B + \beta_1 \frac{L_P^2}{L^2} q_F \ ;\qquad 
q_2 = q_B + \beta_2 \frac{L_P^2}{L^2} q_F  \:, \end{equation}
which gives the dynamical equations as
\begin{equation} \ddot q_1 = - \frac{\alpha^2}{L^2} q_1\ ; \qquad \ddot q_2 = - \frac{\alpha^2}{L^2} q_2 \:. \end{equation}
 
 We can now proceed to write the action in terms of unified variables by first defining 
\begin{equation} {\dot{{Q}}_B} = \frac{1}{L} (i\alpha q_B + L \dot{q}_B); \qquad  {\dot{{Q}}_F}
= \frac{1}{L} (i\alpha q_F + L \dot{q}_F) \end{equation}
and
\begin{equation} \dot{{Q}}_{1} ^\dagger   =   \dot{{Q}}_{B}^\dagger + \dfrac{L_{p}^{2}}{L^{2}} \beta_{1} \dot{{Q}}_{F}^\dagger  ; \ \qquad \dot {{Q}}_{2} =  \dot{{Q}}_{B} + \dfrac{L_{p}^{2}}{L^{2}} \beta_{2} \dot{{Q}}_{F} \:, \end{equation}
which reduces the fundamental action to the following simple and elegant unified form,
\begin{equation} \frac{S}{\hbar} = \frac{1}{2}\int \frac{d\tau}{\tau_{Pl}}\; \text{Tr} \biggl[\biggr. \dfrac{L_{p}^{2}}{L^{2}}  \dot{Q}_{1}^\dagger\;  \dot {Q}_{2} \biggr] \:. 
\label{fundlag}
\end{equation}
This action obeys the unbroken $\text{E}_8\otimes \text{E}_8$ symmetry and is the action for an STM atom prior to left-right chiral symmetry breaking. It has the simple and elegant form of the kinetic energy of a free particle, just as in classical mechanics. Only, the two variables $Q_1$ and $Q_2$ are distinct from each other, which suggests that the STM atom is a two-dimensional entity, as if it were a 2-brane. These two dimensions, along with the eight octonionic dimensions of the underlying space, define a 10D structure evolving in the eleventh dimension, namely Connes time $\tau$. The beauty of this action principle for an atom of apace-time-matter is that it combines spacetime, internal symmetry space, and the overlying fields, into an all-encompassing entity described by two matrices $Q_1, Q_2$ and obeying the $\text{E}_8 \times \text{E}_8$ symmetry.

The action above is for one STM atom, and the action for the universe is a sum over all STM atoms. At the level of the action principle, there is no interaction. But interaction is there in the dynamics in the form of entanglement of states - entanglement is the GTD analog of particle collisions in classical mechanics. Entanglement, followed by spontaneous localization, leads to the classical world, with interactions now mediated by bosonic fields extending over spacetime. It remains to be understood as to why only fermions undergo spontaneous localization but bosons do not. We also need to understand what agency is responsible for enabling entanglement; possibly, the $\text{E}_8 \times  \text{E}_8$ unified force itself might be responsible for entanglement taking place, as evolution progresses.

\subsubsection{\bf Bateman oscillator, and its relation to the STM atom Lagrangian:}
\label{sec:bateman-stm}
The fundamental trace dynamics Lagrangian in Eq.~(\ref{fundlag}) is a special case of the so-called Bateman oscillator\cite{Bateman1931}. We point out the similarities and differences below, with an eye on future development of this unification theory with regard to emergence of the classical limit.

\paragraph{\it Classical Bateman oscillator:}
The Bateman construction embeds a damped oscillator into a conservative system by doubling variables.
With real coordinates $x(t)$ and $y(t)$, damping $\gamma\ge 0$, and spring $k\ge 0$, the
time–independent Lagrangian is
\begin{equation}
L_{\text{Bateman}}
= \dot x\,\dot y
+\frac{\gamma}{2}\!\left(x\,\dot y-\dot x\,y\right)
- k\,x\,y .
\label{eq:BatemanL}
\end{equation}
The Euler–Lagrange equations are
\begin{equation}
\ddot x+\gamma\,\dot x+k\,x=0, \qquad
\ddot y-\gamma\,\dot y+k\,y=0,
\end{equation}
i.e.\ a damped mode and its time–reversed (amplified) partner. With $\gamma=0=k$ one has $L=\dot x\,\dot y$ and $\ddot x=\ddot y=0$.

\paragraph{\it Normal modes and energy signature:}
Define $q_\pm=(x\pm y)/\sqrt2$. For $\gamma=0$,
\begin{equation}
L=\frac{1}{2}\left(\dot q_+^2-\dot q_-^2\right)
-\frac{k}{2}\left(q_+^2-q_-^2\right),
\end{equation}
so both $q_\pm$ satisfy $\ddot q_\pm+k\,q_\pm=0$, but the kinetic form is indefinite (one negative–energy, i.e.\ “ghost”, mode). This indefinite signature is intrinsic to Bateman’s conservative embedding.

\paragraph{\it STM atom in trace dynamics and the free Bateman limit:}
For an STM “atom” evolving in Connes time $\tau$, our basic trace–dynamics Lagrangian is
\begin{equation}
\mathcal L_0 \;=\; \mathrm{Tr}\!\left(\dot Q_1\,\dot Q_2\right),
\label{eq:L0}
\end{equation}
with matrix–valued $Q_{1,2}(\tau)$ and $\dot{\phantom Q} := d/d\tau$.
Varying under $\mathrm{Tr}$ and using cyclicity gives
\begin{equation}
\ddot Q_1 \;=\; 0 \;=\; \ddot Q_2 .
\end{equation}
Thus $\mathcal L_0$ is precisely the \emph{matrix analogue} of the \(\gamma=0,\ k=0\) Bateman limit $L=\dot x\,\dot y$.
As in the scalar case, the kinetic form is cross–coupled (indefinite in normal–mode variables),
but the equations of motion are free and linear.

\paragraph{\it Bateman‑type generalizations for STM atom:}
A conservative Bateman–like extension that introduces a scale and (optionally) a gain–loss structure is
\begin{equation}
\boxed{\quad
\mathcal L
= \mathrm{Tr}\!\left(
\dot Q_1\,\dot Q_2
+\frac{\gamma}{2}\left(Q_1\,\dot Q_2-\dot Q_1\,Q_2\right)
- k\,Q_1\,Q_2
\right),\qquad \gamma,k\in\mathbb R \quad}
\label{eq:STM-Bateman}
\end{equation}
where matrix products are ordinary operator products and the trace ensures reparametrization of factors is immaterial.

\paragraph{\it Equations of motion:}
Varying \eqref{eq:STM-Bateman} and integrating by parts in $\tau$,
\begin{align}
\ddot Q_1+\gamma\,\dot Q_1 + k\,Q_1 &= 0,\\
\ddot Q_2-\gamma\,\dot Q_2 + k\,Q_2 &= 0,
\end{align}
the exact matrix counterparts of Bateman’s scalar equations. Special cases:
\begin{itemize}[leftmargin=1.25em]
\item \(\gamma=0,\ k>0\): \(\ddot Q_i+k\,Q_i=0\) (two harmonic matrix modes). The normal–mode kinetic form remains indefinite (one ghost), but classical solutions are stable.
\item \(\gamma\neq 0\): a gain–loss pair with \emph{conserved} total trace–Hamiltonian (the Bateman mechanism).
\end{itemize}

\paragraph{\it On alternative $k$–terms.}
A “self–quadratic” addition \(\tfrac{k}{2}\mathrm{Tr}(Q_1^2+Q_2^2)\) also yields \(\ddot Q_i+k\,Q_i=0\).
In normal modes it still leaves one negative–kinetic direction; i.e.\ it does not remove the intrinsic Bateman ghost. If a positive–definite kinetic metric is required, a regulator like
\(\epsilon\,\mathrm{Tr}(\dot Q_1^2+\dot Q_2^2)\) (with $\epsilon\!\to\!0^+$ at the end) can be used at the cost of departing from the strict Bateman archetype.

\paragraph{\it Interpretation and guidance:}
\begin{itemize}[leftmargin=1.25em]
\item \textbf{Analogy:} \(\mathcal L_0=\mathrm{Tr}(\dot Q_1\dot Q_2)\) is the free Bateman limit. Equation–of–motion level equivalence is exact.
\item \textbf{Including a scale:} Add the cross “spring” \(-k\,\mathrm{Tr}(Q_1Q_2)\). This preserves the conservative Bateman structure and introduces a natural frequency \(\sqrt{k}\) in Connes time.
\item \textbf{Arrow of time (optional):} Turn on \(\gamma\) in \eqref{eq:STM-Bateman} to realize a gain–loss pair while keeping a time–independent Lagrangian.
\item \textbf{Caveat:} Bateman’s embedding is conservative but energetically indefinite in normal modes (one ghost). This is not a dynamical instability at the classical level, but it matters for quantization and for defining a positive energy. One has to decide explicitly whether the fundamental STM layer tolerates this (as in conservative non-Hermitian / PT symmetric models) or whether a small positive–definite kinetic regulator is desired.
\end{itemize}

In future work, we hope to investigate the
Bateman-like extension of GTD to see if this helps sharpen the comparison with CFS.

\section{Comparison} \label{seccompare}

In this section, we compare and contrast  Causal Fermion Systems, Non-Commutative Geometry, and (Generalized) Trace Dynamics, focusing on their mathematical structures as well as their  physical implications.  

\subsection{Fundamental structures} \label{secfundamental}
We begin the comparison with the fundamental structures. 
\subsubsection{CFS}
The fundamental structure of CFS is the triplet $(\H, \F,\rho)$ where the measure $\rho$ is defined on the set of operators $\F\subset \Lin(\H)$. The measure is the central object studied in the theory. The set~$\F$ is the ambient object within which the physical system is defined. The support of the measure~$\rho$ distinguishes the local correlation operators realized in the physical system under consideration. Once these local correlation operators have been identified, they contain all the information to obtain a spacetime with topological and causal structures as well as wave functions, fields and all other structures needed in physics.
Finally, the measure~$\rho$ also gives an integration measure on the spacetime points.

\subsubsection{NCG} The fundamental structure of NCG is the spectral triple $T=(A,\H,D)$, with the Dirac operator $D$ being the central object. The algebra $A$ (acting on the Hilbert space $\H$) encodes the topology and smooth structure of the underlying space, while $H$ carries the spinor representation. The geometry is tied together by the  Dirac operator $D$, which encodes the metric structure and generates a representation of the differential calculus. Together, $(A,\H,D)$ contain all the information needed to reconstruct the analogue of a spacetime: its topology, differentiable and metric structures, and the integration measure (via traces on $\H$). It is ultimately the spectrum of $D$ that encodes the dynamics of the geometry. Additional axioms constraining the elements of the triple ensure that these data faithfully generalize the notion of a spin manifold.

\subsubsection{GTD}
Generalized trace dynamics postulates that the dynamical variables describing the ``atoms of spacetime'' are given by (Grassmann-number valued) matrices $\{{\bf q}, {\bf p}\}$. These matrix-valued variables are a central object studied in the theory.

One can regard the structure of GTD as a fiber space (a principal bundle) over a base space - the latter being a collection of 16D split bi-octonion-valued and hence operator-valued `points'. The gauge group on the fiber is $\text{E}_8 \times \text{E}_8$. Symmetry breaking (which coincides with the electroweak symmetry breaking) is also a quantum-to-classical transition that results in the splitting of the fiber space into the tangent space and an internal space. The tangent space corresponds to a spacetime 
labeled exclusively by split bi-quaternions (two copies of overlapping 4D spacetimes together constituting a 6D spacetime). The internal fiber space is labeled by the eight additional directions (four + four) belonging to split bi-octonions, over and above the directions already contained in the quaternions. The origin of 4D spacetime (i.e. why there are three directions of space and one of time) appears to lie in the fact that a quaternion has one real and three imaginary directions. The weak force and gravitation are associated with the tangent space, whereas the strong force, electromagnetism, newly predicted $SU(3)$ strong gravity, and a newly predicted $U(1)$ dark electromagnetism are in the internal fiber space. That is, the broken symmetries are in the tangent space and unbroken symmetries are in the internal fiber space (principal bundle).
This elegant separation between broken (external) symmetries and unbroken (internal) symmetries is highly encouraging as to the credibility of the overall construction \cite{SinghST2025}.

\subsubsection{Discussion}
At first glance, the fundamental structures of the three theories seem quite different, however on closer inspection a number of commonalities begin to emerge. To begin with, all three approaches associate to each point in  spacetime a finite-dimensional internal space. In GTD, the internal space is the finite-dimensional Hilbert space on which the operator-valued degrees of freedom \{q, p\} act. For NCG, one has the internal geometry $(A_F,\H_F,D_F)$, while for CFS there is the spin space $S_x$ associated with the spacetime point operator $x\in F$, with dimension at most  $2n$, where $n$ is the spin dimension of the CFS. 

CFS and NCG both feature a Hilbert space explicitly as part of their foundational structures. An intriguing connection is that in CFS, the Hilbert space is constructed from the solutions of the Dirac equation, which inherently encodes details of the Dirac operator. In essence, the measure $\rho$ singles out the physical elements of $\F$, which then encode the content of $D$. In the emergent spacetime the collection of physical wave functions represent the occupied fermionic particle states, so the Hilbert space is directly related to the matter content of our physical reality. In NCG, rather than the solutions of the Dirac equation, one works directly with the Dirac operator as a fundamental object. This is represented, together with the algebra on the abstract  space of square integrable spinors. GTD likewise presupposes a Hilbert space as the representation space on which the operator-valued degrees of freedom act, and with respect to which the spectral action is defined, although it is not usually spelled out as a distinguished input of the formalism. Furthermore, there is 
an explicit Dirac operator built into the formalism. 
GTD begins with generic operator-valued variables \{q,p\}, and through the statistical mechanics of these matrix variables, an emergent quantum field theory appears. In that emergent description, the canonical anti-commutation relations of fermionic degrees of freedom give rise to a Dirac-type dynamics for fermions. 

In both CFS and NCG, bosons and fermions are described on very different footings. In NCG, the fermion species and representations are encoded directly in the choice of Hilbert space, and the representation of the algebra on that Hilbert space. The  bosons on the other hand arise from fluctuations of the Dirac operator $D$. Similarly, the input data of a CFS establishes the physical wave functions directly through its input, these wave functions turn out to be fermionic and satisfy the Pauli exclusion principle~ \cite{fischer2025causal}. The bosonic content then arises from the collective perturbations of this data and hence, since it is not fundamental, its statistic is not subject to the exclusion principle. In contrast with NCG and CFS, GTD treats bosons and fermions within the same operator-theoretic framework from the outset: the fundamental variables \{q,p\} may be bosonic (commuting matrices) or fermionic (Grassmann-valued matrices), but both evolve under the same trace dynamics and are governed by the same conservation laws, while geometry emerges at the statistical level.

\subsection{Emergence of classical spacetime} \label{secemergence}
When we postulate a new theory to replace an already very successful one, the first question to ask is how the previous theory fits into the new picture. The easiest way to answer this question is if one can show that the previous theory emerges as a suitable limit in the new theory. For example, in the case of General Relativity, Newton's gravity emerges in the weak field limit. For approaches generalizing Einstein's theory of gravity, the question is thus how to recover the usual notion of a spacetime from the fundamental mathematical structures and how are they encoded.

\subsubsection{CFS} CFS is a fundamentally relational theory, in which  the set $\F$   encodes causal relations between its elements, and no spacetime manifold is assumed a priori. 
As worked out in~\cite{Finster2021} the subset of regular points in $\F$ (i.e. all operators in $\F$ having maximal rank $2n$) form a, potentially infinite-dimensional, manifold. Moreover, each regular point $x\in \F$ is naturally associated with a finite-dimensional complex vector space, the spin space  $S_x$, of dimension $2n$. This additional structure plays a central role in connecting the fundamental description to fermionic fields in the continuum.  It was shown in~\cite{lqg} that notions of connections and curvature can be defined on the fundamental level of a CFS, without reference to a background spacetime.

The emergence of classical spacetime is addressed through a suitable continuum limit. To this end, one introduces the local correlation map~\eqref{eq:localcorrelation}, which associates points of a classical spacetime with operators in $\F$. In this construction, the measure $\rho$  is supported on the set of local correlation operators corresponding to points in the classical spacetime under consideration. For the local correlation map to be well defined, a regularization is typically required, for instance by discretizing spacetime or introducing an ultraviolet cut-off in momentum space. One then shows that, upon removing the regularization, the resulting causal fermion system is a minimizer of the causal action principle in an well-defined sense. The central objective is to relate the fundamental notions of connection and curvature defined within the CFS framework to their classical counterparts in the limiting spacetime.

One advantage of CFS is that one already starts with a manifold structure, namely the subset of regular operators in $\F$. Although this manifold is not identified a priori with physical spacetime, it provides the geometric arena from which a classical spacetime can emerge in the continuum limit. Identifying the geometric structure is less obvious,  however, we note that CFS satisfies the assumption of the Hawking-King-McCarthy and Malament theorem~\cite{hawking1976new,malament1977class} . As a consequence, the causal relations, supplemented by the volume information encoded in the measure, determine the Lorentzian geometry of the emergent spacetime up to conformal equivalence.
 Within this identification, the spin space $S_x$
 at each regular point can be naturally interpreted as the fiber of a spinor bundle over the classical spacetime.

\subsubsection{NCG} The Standard Model of particle physics naturally contains the elements of a spectral geometry: the known particle content fixes the Hilbert space $\H$, the dynamics selects the ‘Dirac’ operator $D$, and the algebra $A$ can be inferred from the symmetries. 
The spectral triple describing nature consists of two parts $T=T_c\times T_F$. The external spectral triple $T_c$ directly and rigorously encodes the external spacetime. Connes'
reconstruction theorem~\cite{Connes2013SF} allows one to go back and forth between the usual manifold and metric data, and the spectral triple language.
The rules of calculus necessary for defining differentiation, curvature, orientation, and integration in the external space $T_c$ are then borrowed from compact Riemannian spin geometry, and applied to the internal space $T_F$ in order to realize the internal geometry. To be clear: there is no emergence of spacetime within some limit of the construction. The external spacetime is described directly by $T_c$. The novelty is that the rules of this geometry are extended to the internal space $T_F$.

The spectral triple framework is, of course, flexible enough to explore different possibilities for resolving spacetime beyond smooth Riemannian manifolds~\cite{connes-reality}.  At present, however, the NCG 
standard model~\cite{chamseddine-connes-marcolli} is conservative in that it takes the standard Riemannian data for the external spacetime - with the main focus being on understanding the internal geometry of the model. Note that a common criticism of the spectral geometry program is that Connes' reconstruction theorem holds for  compact Riemannian geometries, and not for Lorentzian signature. It should be stressed, however, that NCG does not exclude Lorentzian signature, a rigorous correspondence is simply far more technically demanding in the Lorentzian setting. Considerable effort has been made to extend reconstruction in this direction~\cite{besnard-bizi, Franco_2014, moretti-ncg,strohmaier-ncg, Devastato_2018}. 

\subsubsection{GTD}

The pre-geometric structure of GTD consists of an ensemble of atoms of spacetime-matter. Their internal structure is compatible with a suitably chosen gauge group. As mentioned in Section~\ref{secoverview_TD}, their relation is conjectured to arise dynamically from the entanglement of the different space-time atoms. In the conjectured ``continuum'' limit, the relations between internal spaces go over to a connection of the fiber bundle for which the fiber is isomorphic to the internal space of the spacetime atoms. By identifying a suitable subspace of the fiber space with the tangent space of the continuum manifold, we obtain a connection on the tangent space which is equivalent to having a metric on the tangent space.

\subsubsection{Discussion} 

All three theories agree on the fact that the fundamental structures feature a finite-dimensional internal space. This internal space carries, in one way or another, the information about the gauge fields of the theory. Whether these finite internal structures organize into fiber spaces plays out differently in the three approaches. In CFS, under appropriate continuum-limit constructions, the spin spaces $S_x$
 can be identified with the fibers of a spinor bundle over the emergent spacetime. In NCG the bundle-like organization is present algebraically from the start, as the model is constructed as a product between the external (commutative) geometry and a finite internal geometry, so no limiting procedure is needed. In GTD the passage to bundle structures is conjectural and depends on dynamical assumptions.

NCG and CFS both encode information about continuum geometry in their fundamental mathematical structures, but in very different ways. 
In NCG the external manifold is not something that emerges from a limit: it is built in from the outset through the commutative factor $C^\infty(M)$ in the spectral triple $(\mathcal A,\H,D)$. 
The Dirac operator $D$ is then part of the defining data, tying the manifold structure to spectral information. 
In CFS, by contrast, no background manifold is assumed; instead, geometric information is encoded in the operator relations within $\mathcal F$ together with the measure $\rho$, and in suitable continuum limits this data can reproduce manifold and spinor-bundle structures. 
Thus, while both frameworks give a central role to Dirac-type operators, in NCG the Dirac operator is fundamental input, whereas in CFS Dirac-type dynamics for physical wave functions arise only in the emergent continuum description.

For GTD the assertion is that, in analogy with the Einstein hole argument, classical, macroscopic objects and classical spacetime emerge side by side.
 In this framework, both geometry and classical matter arise as effective descriptions obtained from an underlying non-commutative dynamics.
For CFS, this is not the case. While a classical spacetime geometry emerges in an appropriate continuum limit, the fundamental dynamical objects remain fermionic wave functions governed by the Euler Lagrange equation of the causal action principle. The emergence of spacetime in CFS therefore does not coincide with a classical limit of the matter degrees of freedom.
 As discussed in~\cite{fischer2025causal}, for a CFS, the classical spacetime emerges in the limit where the dimension of the Hilbert space becomes large. Hence, the emergent space is classical due to an averaging over many degrees of freedom.  

A key structural difference between GTD and both NCG and CFS, is that at the fundamental level  there are no obvious a priori relations between distinct “atoms of spacetime–matter.” In NCG, the continuous geometry $T_c$ encodes spatial relations between the internal spaces. Similarly, in CFS all elements  $x,y\in \F$  act as operators on the same Hilbert space, and their mutual relations are encoded directly in their operator product. In this sense, relational structure is built into the CFS framework from the outset, rather than arising only through interactions. 
For GTD it remains an important open challenge to understand how relations between different spacetime atoms arise. Since entanglement is generally expected to emerge through interactions, this naturally motivates a comparison of the respective action principles underlying the different approaches.

\subsection{Appearance of Quantum Fields}\label{secquantization} A cornerstone of modern physics is the fact that matter around us is governed by quantum theory. This has lead to the expectation that any fundamental theory needs to be quantum and include a quantization of gravity. The three candidates discussed in this paper are not straightforward generalizations of conventional quantization procedures. Accordingly the question is, whether and to what degree each of them is ``quantum''. 

\subsubsection{CFS} On the fundamental level, causal fermion systems are described by self-adjoint operators on a Hilbert space of wave functions. These are the exact building blocks of quantum mechanics, with slightly different roles than usual. More specifically, in the continuum limit description, the wave equations satisfy the Dirac equation. Together with the collapse model derived in~\cite{finster2024causal} and the position observables defined in~\cite{fischer2025causal}, this gives a complete model of first quantization. Beyond that, Quantum field theory including second-quantized fermionic and bosonic fields are obtained in a different limiting case worked out in~\cite{fockfermionic, fockentangle, fockdynamics}.
Whether and to what degree this framework qualifies as a theory of quantum gravity is not clear at present as the standard Lorentzian manifold structure is replaced by something more general in the fundamental description. 

\subsubsection{NCG} Once the spectral triple of the NCG SM is established, one obtains dynamics via the spectral action principle. The heat kernel expansion then gives the familiar standard model action coupled to Einstein Hilbert gravity plus higher order correction terms. The standard approach is to interpret this action as a classical action, to be quantized following all of the usual techniques. In other words, the standard conservative approach to the NCG SM is to perform quantization \emph{after} the heat kernel expansion, rather than before, and as such the approach  has very little new to say about the quantization of geometry.

This, however, does not mean that NCG is entirely silent on the issue of quantization. In the NCG approach to gauge theory the bosonic data is entirely unified as  fluctuations of the Dirac operator $D$. In particular, the metric encodes information about distances in the external spacetime, while the Higgs encodes information about distances in the discrete internal space (gauge bosons encode `mixed' metric data). While we don't yet have a direct experimental handle on quantized gravitons, we do have an experimental handle on gauge and Higgs fields. This means that within the NCG framework, these fields  provide an experimentally accessible window into what it actually means to quantize a geometry.

The discrete internal geometry of the NCG SM is a two point space, corresponding to the two chiral sectors of the standard model. The chiral sectors of the model are coupled by the Higgs field, which is encoded in off-diagonal elements in the finite Dirac operator $D_F$ of the internal space $T_F$. That is, the internal space is entirely non-dynamical, except for the dynamics of the spacing between the discrete points (i.e. chiral sectors). A number of authors \cite{rov99,besquan,hale}
have looked specifically at isolating this discrete internal space in toy models, and 
quantizing the dynamics. What emerges, is a tantalizing view of discrete, quantized, geometry where the internal points  exist at discrete quantized distances from one another.

\subsubsection{GTD} In GTD the act of quantization consists of raising the classical real-number-valued dynamical variables to the status of matrices. This converts classical dynamics to a pre-quantum theory, because of the newly arising Adler-Millard charge \eqref{eq:adler-miller-charge}. From this pre-quantum dynamics, quantum field theory is emergent in a statistical thermodynamic approximation, upon coarse-graining over the time scales over which the dynamics is observed.

\subsubsection{Discussion}

CFS and GTD agree in the context of QFT that it is an emergent theory. They differ however in the context of first quantization which is essentially baked into the construction of CFS. 
For CFS a smoking gun of its quantum nature is that when regularizing the local correlation map in the continuum limit, the theory has a minimum length scale with a non-zero self-separation. This is considered by some, e.g.~\cite{padmanabhan2022microscopic,pesci2019spacetime} as a tell tale sign of quantum gravity. Since the gravitational coupling scales with the regularization scale, one could conclude, that the fact that spacetime is generally not flat is actually an experimental signature of quantum gravity. 
For NCG this signature isn't as generic, as in the conservative approach the NCG SM, the points of the continuous spacetime factor $T_C$  are not separated by any minimal length; the underlying manifold remains a classical continuum. A spectral triple specifies, in a precise sense, the kinematic configuration space of the model: the algebra encodes generalized coordinates, the Hilbert space carries the representation of states, and the Dirac operator determines the geometric data. The spectral action then defines the dynamics. In practice, one evaluates this action through a heat kernel expansion, obtaining an effective classical Lagrangian, after which the theory is quantized in the usual field-theoretic manner. 

Of course, one need not restrict attention to such conservative models. The framework is rich enough that one could in principle replace the external spacetime with something that is more inherently quantum. This could include $\theta$-deformed $\mathbb{R}^4$~\cite{10.1143/PTPS.171.11}, $\kappa$-Minkowski~\cite{MATASSA2014136}, large N discrete spacetimes~\cite{universe8040215}, Spectral triples with compact resolvent but no underlying manifold~\cite{gabriel2013ergodicactionsspectraltriples}, coarse-grained geometries based on operator systems~\cite{Walter22}, and more.  Each of these approaches modifies the notion of spacetime at a structural level, rather than merely introducing quantization after the fact.

Moreover, one may contemplate quantizing directly at the level of the spectral action itself, rather than only after performing the heat kernel expansion and reducing to an effective classical field theory~\cite{rov99,besquan,hale}. In this sense, non-commutative geometry does not exclude quantized or fundamentally non-classical spacetime structures. Rather, it provides a flexible mathematical framework within which such possibilities can be explored, although it remains an open question which of these directions, if any, will ultimately prove physically viable.

Comparing CFS and NCG an interesting aspect is that they both give rise to position observables. In NCG they arise from the coordinate algebras, while in CFS \cite{fischer2025causal} they are defined via the projection operators $\pi_x$ on the spin-spaces associated with the spacetime point operators $X$. In an upcoming paper\cite{yadav_heisenberg} some of the present authors show, that for stationary CFS, i.e. measures with a unitary family $U_t$ of symmetries $\rho(x)=\rho(U_t x U_t^{-1})$ these observables satisfy the canonical Heisenberg evolution equation.  

Finally, the most appropriate way to classify GTD is to say that it is a pre-quantum, 
pre-spacetime theory. It has the same dynamical construction as that of classical dynamics and classical spacetime: only, real number-valued variables and spacetime points are replaced by operators/matrices and non-commutative/non-associative structures. Quantum theory sits half-way between GTD and classical dynamics.

\subsection{Action Principle}\label{secaction}
A common feature of most modern physical theories is that they outline an action principle which chooses from the set of admissible mathematical structures those that correspond to physical systems. For general relativity, for example, the set of admissible mathematical structures is the set of all orientable Lorentzian manifolds, while the physical ones are only those minimizing the Einstein-Hilbert action. 
\subsubsection{CFS}
The Lagrangian in the causal action principle is built from the eigenvalues of the closed chain of the generalized two-point correlator (see, e.g., the discussion in~\cite{fischer2025causal})). The Lagrangian is compatible with the causal structure of the theory in the sense that spacelike separated points do not contribute to the action. In the case of Minkowski space, the kernel of the two-point correlator is built from the solutions of the Dirac equations and thus itself solves the Dirac equation. The action is non-local in the sense that it is built from the two-point correlations across spacetime. However, in important examples, the action is concentrated near the diagonal of $\F \times \F$, i.e. the Lagrangian is of short range and falls of quickly. As a result most of the action comes from points $y\approx x$. 

\subsubsection{NCG}
The Lagrangian in Connes spectral action is the trace of the Dirac operator. In the Riemannian setting it is known that the spectrum of the Dirac operator encodes the full geometric information.  The spectral action is fundamentally nonlocal, as the spectrum of the Dirac operator encodes the global geometry. 

\subsubsection{GTD}
The gravitational Lagrangian in GTD is motivated by the spectral action principle of NCG, and given by the trace of a suitable product of 
momentum variable matrices. The idea is that the original spectral action is expressed in terms of Dirac eigenvalues; in GTD each eigenvalue is raised to the status of an operator. This operator is the very Dirac operator whose eigenvalues are being used. 

\subsubsection{Discussion}
A key commonality
between the CFS and NCG actions 
is that they 
are both nonlocal, albeit in a different manner. In CFS the non-locality is explicit: the action is built from a two-point Lagrangian depending on pairs of spacetime points. In NCG, by contrast, nonlocality enters more subtly through spectral data. As for GTD, since it builds on the spectral principle of NCG, it is also non-local. It should be noted though that at this stage it is not clear how interactions between STM atoms arise in GTD: are interactions already present (via entanglement) in the currently proposed GTD Lagrangian, or the Lagrangian needs to be generalized to bring in interactions? 

Although both NCG and CFS encode physics through the Dirac operator, they do so in different manners. In NCG, the central object is the spectrum of the Dirac operator itself, and the spectral action depends only on its eigenvalues.
 In CFS, by contrast, one considers a two-point correlation operator constructed from a subset of solutions of the Dirac equation. The set $\F$ is then the set of all such correlation operators, and the  measure $\rho$,  as the central dynamical object of the theory, encodes which correlations are realized in a given physical system. 
 
 At present, it is unclear whether these are just different ways to encode the same information or whether they actually encode different aspects of the information contained in the Dirac operator. 

Part of the difficulty in addressing this question lies in the fact that the CFS framework is intrinsically Lorentzian. This is reflected in the causal action principle, which is tightly linked to the causal structure of spacetime. 
By contrast, a fully satisfactory Lorentzian formulation of the spectral action principle in NCG is still missing.
The main obstruction is that the spectral action is defined via a regularized trace of the Dirac operator, specifically, the Dixmier trace \cite{Chamseddine_1997}. This trace is well-defined only in the Riemannian setting.
 In Lorentzian signature, however, the Dirac operator is not symmetric on a Hilbert space but instead acts on an indefinite inner product space, meaning that the standard functional-analytic tools no longer apply.

Putting these conceptual differences aside, there are nevertheless striking technical similarities.
 To compute the spectral action via the Dixmier trace, one studies the asymptotic behavior of large eigenvalues, which is encoded in the behavior of the resolvent near the diagonal.
The corresponding heat kernel expansion can be regarded as
the elliptic analog of the light-cone expansion~\eqref{lce} (or, more generally, the Hadamard expansion) used to analyze the causal action in the continuum limit. With this in mind, the causal action principle can even be regarded as the Lorentzian analog of Connes' spectral action principle.
This analogy can be worked out in more detail for static causal fermion systems, because in this case
the causal action principle can be analyzed with elliptic methods. We plan to study these connections
separately in~\cite{position}.

Among the three theories, GTD has the most general action principle, as it isn't so much a single Lagrangian, rather than a functional type of Lagrangian. Connes' spectral action in the context of NCG is clearly a special case of this family of Lagrangian as it is of trace class; however, generically an infinite-dimensional trace. For the causal action principle, the situation is more subtle as we now elaborate.  

The causal action, being exclusively relational, connects different spacetime points. 
The causal Lagrangian~\eqref{eq:lagrangian} therefore 
depends on two arguments~$x$ and~$y$. It is clear that generically the causal Lagrangian is not polynomial due to the appearance of the absolute values of the eigenvalues. 
However, as already mentioned, in important examples such as Minkowski space, the Lagrangian is concentrated near the diagonal in $\F\times\F$, that is, the main contribution to the causal action is obtained when the spacetimes points~$x$ and~$y$ are close together.
This motivates the consideration of the limiting case of the causal action
in which the Lagrangian is evaluated only on the diagonal~$x=y$,
\beq \label{Sdiag}
\text{\em{diagonal action:}} \qquad \Sact^\text{diag}(\rho) := \int_\F \L(x,x)\: d\rho(x) \:.
\eeq

\begin{Lemma} For a CFS with spin dimension~$n$, the causal Lagrangian
evaluated on the diagonal can be written as
\beq \label{Ldiag}
\L(x,x) = \tr\big( x^4 \big) -\frac{1}{2n} \tr\big( x^2\big)^2 \:.
\eeq
If the spacetime point operator~$x$ is regular (in the sense that its rank is~$2n$), we have
\beq \label{Ldiag2}
\L(x,x) = \tr\big( Y(x)^2 \big) \qquad \text{with} \qquad Y(x) := x^2 - \frac{1}{2n} \tr(x^2)\: \pi_x \:,
\eeq
where~$\pi_x$ is the orthogonal projection operator onto the image of~$x$.
\end{Lemma}
\Proof Let~$x \in \F$. We denote its non-trivial eigenvalues by
\begin{equation} \nu_1 \leq \cdots \leq \nu_n \leq 0 \leq \nu_{n_1} \leq \cdots \leq \nu_{2n} \:. \end{equation}
Then
\begin{align*}
\L(x,x) &= \frac{1}{4n} \sum_{i,j=1}^{2n} \big( \nu^2_i - \nu^2_j \big)^2
= \frac{1}{4n} \sum_{i,j=1}^{2n} \big( \nu^4_i - 2 \nu^2_i \nu^2_j + \nu^4_j \big) \\
&= \sum_{i=1}^{2n} \nu^4_i - \frac{1}{2n} \sum_{i,j=1}^{2n} \nu^2_i \nu^2_j \:
= \tr\big( x^4 \big) -\frac{1}{2n} \tr\big( x^2\big)^2 \:,
\end{align*}
proving~\eqref{Ldiag}. In order to verify~\eqref{Ldiag2}, we expand out to obtain
\begin{align*}
\tr\big( Y^2 \big) &= \tr\big( x^4 \big) - \frac{1}{n} \tr \big( x^2 \big) \tr \big( \pi_x\: x^2 \big)
+ \frac{1}{4n^2} \tr(x^2)^2\: \tr \big(\pi_x \big) \\
&= \tr\big( x^4 \big) - \frac{1}{n} \tr \big( x^2 \big)^2
+ \frac{1}{2n} \tr(x^2)^2 = \tr\big( x^4 \big) -\frac{1}{2n} \tr\big( x^2\big)^2 \:.
\end{align*}
This concludes the proof.
\QED

Using~\eqref{Ldiag2} in~\eqref{Sdiag}, we obtain
\begin{equation} \Sact^\text{diag}(\rho) := \int_\F \tr\big( Y(x)^2 \big)\: d\rho(x) \:. \end{equation}
This is a special case of the action of trace dynamics for an ensemble of matrices~$(Y(x))_{x \in M}$.
Clearly, in the description by an ensemble of matrices, spacetime becomes merely an index set
for the matrices. The relations between spacetime points as induced by the causal action principle
do not enter the description. This is consistent with the limiting case~\eqref{Sdiag} of omitting the contributions for~$x \neq y$ in the causal action.

These considerations give an indication that trace dynamics is suitable for describing
the {\em{vacuum configuration}} of a CFS in Minkowski space. It was shown in~\cite{fischer2025causal} that, in this special case, indeed also the full non-local action can be written as a trace. 
With this in mind, the use of trace dynamics in~\cite{singh-octo2}, 
giving rise to the derivation of coupling constants and mass ratios in~\cite{singh-octo1, BhattEtAl2022MajoranaEJA, singh2025fermionmassratiosexceptional},
may carry over to similar results for the causal action principle.
However, this adaptation does not seem to be straightforward.

Now we mentioned in the previous section on the emergence of the spacetime continuum that for GTD the spatial relations emerge dynamically through entanglement. However, that requires the presence of an interaction between the different atoms of spacetime. This interaction needs to be represented in the action principle, and  it is unclear how this is encoded in the current setup. For CFS we see that the Lagrangian on the diagonal reduces to an integral over the free ``atoms of spacetime''. The off-diagonal contributions of the Lagrangian in CFS, on the other hand, can  be understood as interaction terms between the different spacetime operators. Given the fact that entanglement corresponds to a specific pattern of correlations, it is intriguing to think of the relational nature of the CFS Lagrangian  to capture the entanglement between different regions in spacetime in the off-diagonal contributions. In summary, we can say that the GTD action is formulated for spacetime atoms, while the CFS action is formulated in terms of relations between spacetime points. One take-away lesson from this comparison is that either there is a hitherto undiscovered mechanism/force (e.g., pre-gravitation) in the GTD action presented here, which results in
entanglement as evolution progresses, or eventually one might have to add off-diagonal/interaction terms to the GTD Lagrangian. 

\subsection{Conservation laws}\label{secconserv}
Conservation laws have been at the heart of physics ever since Noether's famous proof that every symmetry of the action corresponds to a conserved quantity of the dynamical system satisfying the Euler-Lagrange equations. 

\subsubsection{CFS}
The causal action is unitarily invariant, as can easily be seen. The corresponding conserved quantity
is the {\em{commutator inner product}} taking the form (for more details, see~\cite[Section~9.4]{Finster2024}
or~\cite[Section~5]{noether} and~\cite[Section~3]{finster2021linear})
\begin{equation}\label{eq:CFSconservation}
    \la \psi^u | \phi^v \ra^\Omega := -2i \,\bigg( \int_{\Omega} \!d\rho(x) \int_{M \setminus \Omega} \!\!\!\!\!\!\!d\rho(y) 
- \int_{M \setminus \Omega} \!\!\!\!\!\!\!d\rho(x) \int_{\Omega} \!d\rho(y) \bigg)\:
\Sl \psi^u(x) \:|\: Q(x,y)\, \phi^v(y) \Sr_x \:,
\end{equation}  
where~$\Omega$ is the past of a Cauchy surface and~$\psi^u, \phi^v$ are physical wave functions
corresponding to Hilbert space vectors~$u, v \in \H$ (see~\eqref{psirep}).
\subsubsection{NCG}

The spectral action is invariant under the automorphisms of a spectral triple. With respect to these symmetries, the classical Noether correspondence continues to apply: continuous symmetries of the spectral action give rise to conserved currents and charges. This mirrors the familiar Noether link between symmetries and conservation laws in classical field theory, but reformulated in the operator-algebraic language of spectral triples and their automorphism groups. In this sense, the conservation laws obtained in non-commutative geometry parallel the conventional ones, while being derived from the unified geometric framework provided by the spectral triple.

At present, we are not aware of any conservation law that is intrinsically specific to non-commutative geometry. However, since the spectral action is formulated as a trace, arguments analogous to those in generalized trace dynamics (GTD), explained below,  may apply in this setting as well. One potentially relevant observation is that non-commutative geometry admits an intrinsic notion of time evolution, known as Connes time, which arises from the Tomita–Takesaki modular theory of von Neumann algebras~\cite{connes}. This raises the possibility that, within GTD, phase-space evolution along trajectories preserving the Adler–Millard charge may be identified with evolution in Connes time.

\subsubsection{GTD}
As explained in Section~\ref{secoverview_TD}, in trace dynamics, the global unitary invariance of the trace Lagrangian gives rise to the conserved
{\em{Adler-Millard charge}}~$\tilde{C}$ given by (for details see~\cite[Section~2.2]{adler-trace})
\begin{equation} 
\tilde{C} = \sum_{r \in B} [q_r, p_r] - \sum_{r \in F} \{q_r, p_r \} \:. \end{equation}

\subsubsection{Discussion}

Given that both originate from the unitary invariance of the respective action principles, there is a clear similarity between the conservation laws in trace dynamics and the causal action principle, as we now explain.

If we start with the commutator inner product of CFS \eqref{eq:CFSconservation} and represent it in terms of the Hilbert space scalar product by
\beq \label{Sigdef}
\la \psi^u | \phi^v \ra^\Omega = \la u | \Sig\, v \ra_\H \:,
\eeq
we obtain a symmetric operator~$\Sig$ on~$\H$ which is time independent
in the sense that it does not depend on the choice of the set~$\Omega$.
The operator~$\Sig$ can be regarded as a CFS analogue of the operator~$\tilde{C}$ in the theory of GDT. The commutator inner product plays an important role in many applications, e.g., the derivation of collapse models from CFS which we will discuss below in Section \ref{seccollapse}. 

In trace dynamics, it is then argued that, applying the law of equipartition, in the statistical average
the Adler-Millard charge~$\tilde{C}$ should be a multiple of the identity operator. Under natural assumptions
and approximations, this leads to the canonical commutation relations of quantum theory

\beq \label{ccr}
\big[ q_{i, \text{eff}}, p_{j, \text{eff}} \big] = i\,\hbar\, \delta_{ij} \:.
\eeq
(for details see~\cite[Chapter~5]{adler-trace}).
Applying a similar reasoning to the commutator inner product of CFS, one
concludes that also the operator~$\Sig$ should be a multiple of the identity.
This gives a justification for the following assumption.
\begin{Def} \label{defSLrep}
The commutator inner product is said to {\bf{represent the scalar product}} if
\begin{equation} 
\la \psi^u | \psi^v \ra^\Omega = c\, \la u|v \ra_\H \qquad \text{for all~$u,v \in \H^\fermi$} \end{equation}
for a positive constant~$c$ and a subspace~$\H^\fermi \subset \H$.
\end{Def} \noindent
Clearly, in view of the representation~\eqref{Sigdef}, this property can be stated equivalently as
\beq \label{Sigrel}
\Sig|_{\H^\fermi} = c\, \1|_{\H^\fermi} \:.
\eeq
In previous works on CFS (see, for example,~\cite[Definition~3.9]{finster2021linear}),
this property was introduced and used as an
ad hoc assumption motivated by the correspondence to the scalar product on
Dirac wave functions in Minkowski space as established in~\cite[Section~5]{noether}.
The theorem of equipartition gives a more fundamental explanation for this assumption.

We finally explain the role of the subspace~$\H^\fermi$. Taking the trace of~\eqref{ccr},
one readily sees that, in the finite-dimensional setting, this relation cannot hold on the whole
Hilbert space. Similarly, the relation~\eqref{Sigrel} cannot hold if we choose~$\H^\fermi = \H$
(for details see~\cite[Appendix~A]{current}).
Therefore, it is crucial that these relations hold only on a proper subspace of~$\H$.
The idea is that this subspace should contain all the ``low-energy'' wave functions that are
accessible to experiments. Therefore, the identity holds for all practical purposes when describing the world around us.

\subsection{Collapse} \label{seccollapse}
Physical collapse has been proposed in a number of variations
as a possible resolution of the measurement problem in quantum theory
(see~\cite{diosi1, diosi2, pearle0, ghirardi2, ghirardi-pearle-rimini, gisin,
penrose-collapse, penrose-collapse2} or the survey papers~\cite{bassi-ghirardi, ghirardi, pearle}).

\subsubsection{CFS}
In~\cite{finster2024causal} an effective collapse theory for CFS was derived. The effect relies on a plethora of undetectable stochastic fields. The deterministic unitary equations of quantum mechanics are thus to be understood for an ensemble of systems where the probe is initially prepared in an identical manner, but where the collapse onto an eigenspace of the observable occurs in each individual system due to the interaction with the stochastic fields. In a follow-up paper~\cite{finsterNoHeating} it was recently shown that due to the back-reaction with the background via the stochastic fields, this collapse model does not lead to a heating of the probe. The CFS based collapse model is therefore not subject to the experimental constraints in~\cite{bassi, donadi-piscicchia, piscicchia2}.

A recent proposal~\cite{fischer2025causal} suggests that the only fundamental observables of CFS are spacetime position observables. In this picture, collapse would then localize a state in a certain spacetime region. Therefore, the phenomenon of collapse is intimately connected with the emergence of localized macroscopic bodies. 
\subsubsection{NCG}
At present, the authors are not aware of any direct connection between non-commutative geometry and physical collapse models.

\subsubsection{GTD}
In GTD collapse plays an essential role in the emergence of the classical world. Recall that the spacetime atoms were obtained by replacing the eigenvalues of the Dirac operator with matrix-valued dynamical variables. The conjecture is that it takes objective physical collapse for an atom of spacetime to return a unique value. These values then give rise to an effective classical geometry hand in hand with the emergence of classical bodies. 
\subsubsection{Discussion}

Collapse plays a fundamental role in both CFS and GTD in the emergence of classical reality. However, the technical development in CFS is further advanced and, hence, a detailed comparison is not possible at this stage. However, we can remark on a few conceptual points. As already mentioned, the emergence of classical spacetime in CFS is not predicated on physical collapse, in contrast to the conjectured mechanism for GDT. However, the emergence of classical material bodies very much is, as the collapse of subsystems inevitably leads to the collapse of the corresponding macroscopic object, leading to a localization in spacetime of said macroscopic physical body.

\subsection{Derivation of the Standard Model}\label{secSM}
A unifying theory must not only give rise to spacetime and its geometry in a suitable limit, but it must also account for the particles observed in our physical world. Here we compare the various constructions schematically. For CFS and a subset of NCG, i.e., octonion-based theories, this comparison has already been carried out in more detail in~\cite{oct-cfs}.

\subsubsection{CFS}
To construct the Standard model within the CFS framework, one begins by assuming that the vacuum is formed of eight fermionic sectors (one for every quark color and isospin, one for the leptons and one for the neutrinos). This is the necessary input in order to arrive at the correct particle content in the model. From this starting point one can derive the standard model gauge fields as emerging phenomena. In the framework of CFS, each sector automatically contains the anti-particles as well. Breaking the chiral symmetry in the neutrino sector, the framework can model the Standard Model plus sterile right-handed neutrinos. 

\subsubsection{NCG}
 In order to construct gauge theories within the framework of NCG, 
 one begins by prescribing the entire fermionic content of the model in mind. For the standard model, this would be three generations of Up and Down quarks, electrons and neutrinos, left and right handed chiralities, particles and anti-particles. So 32 fermions in total per generation, giving rise to the input Hilbert space $\H_F = \mathbb{C}^{96}$~\cite{chamseddine-connes-marcolli,chamseddine-connes2, boyle-farnsworth}. One then searches for a finite dimensional algebra, with a representation on $\H_F$, which reproduces the desired  gauge symmetries and representations. For the standard model gauge symmetries there are very few available options, and one takes canonically $A_F = \mathbb{C}\oplus \mathbb{H}\oplus M_3(\mathbb{C})$, together with an appropriate representation on $\H_F$.  Given the data $(A,\H_F)$, one is then strongly constrained by the rules of the geometry in the choice of $D_F$~\cite{boyle-farnsworth}. Once the data $T_F=(A_F,D_F,\H_F)$ is selected, one forms a product geometry $T = T_c\times T_F$ with an appropriate external spacetime $T_c$. Fluctuation of the total Dirac operator then produces Gauge and Higgs fields. It is the fluctuated Dirac operator, which is then used to construct dynamics via the spectral action and fermionic terms.

\subsubsection{GTD}
In GTD the particle content of the theory is encoded in the internal structure of the matrix-valued dynamic variables $\{{\bf q}, {\bf p}\}$. The choice of the space in which ${\bf q}$ and ${\bf p}$ take values then leads to a certain particle content in the theory with the bosonic and fermionic content appearing on a similar footing. The particle content is dictated by the chosen symmetry breaking of the $\text{E}_8 \times \text{E}_8$ theory, and three generations of standard model chiral fermions are recovered, along with three types of right-handed sterile neutrinos \cite{kaushik_2024, singh2025fermionmassratiosexceptional}.

\subsubsection{Discussion}
The action principle of GTD, and the scheme for incorporating gravitation in TD is entirely motivated by the spectral action principle of NCG. The key point is that the Dirac eigenvalues from NCG are raised to the status of operators (this being akin to quantization) in GTD. 
Consequently, GTD is a pre-quantum, pre-spacetime theory; therefore, gravitation is already `quantized' because of the way TD works.
 The use of Connes' time in GTD is also motivated by NCG. The concept of internal symmetries as resulting from a non-commutative space is also common to NCG and GTD.
In contrast to the standard approach to NCG, however, the spacetime geometry is made not just non-commutative, but also non-associative.

An interesting observation when comparing GTD and CFS is that both arrive at three generations of fermions, albeit following very different routes. For GTD they stem from the breaking of a large symmetry group. Meanwhile, for CFS they arise in the derivation of the dynamical equations of the bosonic fields as a necessary condition for the initial value problem to be well-posed. Furthermore, both approaches are able to derive the Standard Model with an additional sector of sterile right-handed neutrinos. 

A comparison between the NCG SM and CFS suggests the possibility of a structural connection between the two approaches. In CFS, the fundamental object is the fermionic projector $P_m(x,y)$, which encodes both geometric and matter content through its kernel. In the NCG formulation of the Standard Model, the particle content and gauge symmetries are determined by the finite algebra $A_F$ and its representation on a finite Hilbert space  $H_F$, and the associated finite Dirac operator $D_F$. 
To incorporate internal degrees of freedom into the CFS setting, we enlarge the Hilbert space underlying the fermionic projector. Rather than working with a single Dirac sector, we consider
\begin{align}
    H = L^2(M,S)\otimes H_F
\end{align}
where $L^2(M,S)$ is the space of square integrable spinors, and $H_F$ carries the finite-dimensional representation of the internal algebra. This tensor product structure allows spacetime spin degrees of freedom and internal quantum numbers to coexist in a single Hilbert space, analogous to the separation of external and internal symmetries in the NCG framework. A natural vacuum ansatz for the vacuum projector is then
\begin{align}
    P(x,y) = P_m(x,y)\otimes \mathbb{I}_F,
\end{align}
where $P_m(x,y)$ is the kernel of the vacuum fermionic projector of mass $m$, and $\mathbb{I}_F$ acts trivially on the internal space. This ansatz is conceptually useful but physically incomplete: not all fermions share the same mass, and in the Standard Model the algebra $A_F$ acts differently on left- and right-handed fields. A more realistic construction would require internal projectors distinguishing these sectors and representations that reflect chiral structure.

Even with these caveats, the tensor product structure implies that the spin dimension of the enlarged CFS increases proportionally to 
dim $H_F$, thereby accommodating internal multiplicities analogous to those in the NCG description of the Standard Model. For 
dim $ H = 32$, the resulting spin space dimension coincides with that of the finite Hilbert space used in the Standard Model spectral triple.

Internal symmetries are introduced through the representation of $A_F$ on $H_F$. A central question is the relationship between the Dirac operator $D$ and the fermionic projector $P$. Without regularization, the projector is an apriori unbounded operator that, in Minkowski space, selects the subspace of negative-frequency solutions (the “Dirac sea”). Formally, this is expressed via spectral calculus as
\begin{align}
    P \sim \chi(-\infty,0)(D),  
\end{align}
where $\chi(-\infty,0)(D)$ denotes the characteristic function of the negative spectrum of $D$. This should be understood as a spectral projection rather than an identification of $P$ with the unbounded Dirac operator itself. In Minkowski space the spectral theorem does not apply in the same way as in the compact Euclidean case, so this expression should be understood heuristically.

Deformations of the vacuum projector then encode gauge interactions analogously to how the  inner fluctuations of the Dirac operator generate gauge fields in NCG. A natural ansatz is then that the deformed fermionic projector, should be related to the fluctuated Dirac operator
\begin{align}
    P_A= \sim \chi(-\infty,0)(D_A), 
\end{align}
with modifications acting on the internal tensor factor. 
However, this relation remains heuristic unless a precise map between the functional calculus of fluctuated Dirac operators $D_A$ and deformed projectors $P_A$ can be established.

A central open question is whether internal geometries arising from different finite spectral triples yield minimizers of the causal action principle. In other words, one may ask which choices of internal algebra and representation (as used in NCG particle models) are dynamically compatible with the CFS variational principle and which are excluded. Addressing this question would require a detailed study of how the causal Lagrangian depends on the internal tensor structure and on admissible perturbations of the vacuum projector. Making this correspondence precise would be essential for a consistent unification of the two frameworks.

Now we want to reflect on what is probably the key lesson from CFS for the construction of a fundamental theory: In classical differential geometry the connection encodes how the fiber spaces at neighboring points are related. Through parallel transport this is then extended to a map relating the fiber spaces in a convex neighborhood in a unique manner. 
Now, if we want to be able to relax the smoothness assumption of the underlying spacetime we need a different way to identify how the internal spaces at different spacetime points are related to each other. The way this is achieved in CFS is through a family of wave functions. The core idea is that different spacetime points are linked by the information carried by wave functions that are supported at both points. Formulating this more rigorously, in CFS every spacetime point operator $x$ comes with with an associated spin space $S_x$, where $S_x$ is the image of $x$ in the fundamental Hilbert space $\H$. 

In order to link this picture to NCG, it is useful to define an equivalence relation on $\F$ by $x\sim y$ if $S_x=S_y$. Hence, the equivalence relation identifies spacetime points that share the same local Hilbert space. In addition, note that the operator $x$ defines an indefinite inner product on $S_x$ given by
\begin{equation} \Sl u | v \Sr_x = -\la u | x v \ra_\H \qquad \text{(for all $u,v \in S_xM$)}\:. \end{equation}
and the generalized two-point correlator can then be written as 
\begin{equation} P(x,y) := \sum_{i = 1}^N |{\psi^{u_i}(x)}\Sr \Sl\psi^{u_i}(y)| \:. \end{equation}
where $\psi^{u_i}(x)$ are the physical wave functions defined in \eqref{psirep} associated with the basis vectors $u_i$ of the fundamental Hilbert space. The CFS Lagrangian can then be formulated in terms of this two-point correlator. 

Let us try to take this insight and apply it to NCG. For this, we need to drop the spacetime factor $T_c$ in the NCG construction \eqref{eq:NCGsplit} and replace it with something else. For our construction, we therefore only start with the spectral triple of the internal space $T_F= (A_F, \H_F, D_F)$. Identifying $\H_F$ as the local Hilbert space suggests considering a larger fundamental Hilbert space $\H$. We here restrict our attention to spectral triples where the local Hilbert space is finite- and even-dimensional, $\dim(\H_F)=2n$.
We consider the set of all possible embeddings $\Xi: \{\xi_i: \H_{F}\hookrightarrow\H_{aux}\}$ and define the equivalence relation $\xi_a\sim \xi_b$ by $\xi_a(\H_F)= \xi_b(\H_F)\subset \H$ being the same subspaces of the fundamental Hilbert space. Hence, the two embeddings only differ by a unitary transformation acting on the internal Hilbert space alone. Restricting attention to the quotient space $\Tilde{\Xi}:= \Xi/\sim$ gives a clear separation of the degrees of freedom intrinsic to the construction of the triplet $T_F$ from the internal algebra $A_F$ and the degrees of freedom that arise from the embedding into the fundamental Hilbert space $\H$. 

For this construction to align with CFS the key question concerns the spectrum of the Dirac operator $D_F$ on the internal space. 
\begin{Lemma}\label{lem:nn}
   Let $\dim\H_F=2n$.  If $D_F$ is self-adjoint and anti-commutes with a grading operator, then it has exactly $n$ positive and $n$ negative eigenvalues. The spectrum is symmetric with respect to zero. Moreover, $D_F$ is traceless. 
\end{Lemma}

\Proof

First we note, that $D_F$ being self-adjoint and anti-commuting with the grading operator $\sigma:=\diag (\mathbf{1},-\mathbf{1})$ implies that it is of box-off-diagonal form 
\begin{equation} \{\sigma,D_F\}  =0  \qquad \Rightarrow \qquad D_F =\begin{pmatrix}
0 & X \\ X^* & 0 \end{pmatrix}\:, \end{equation}
where $X^*$ is the adjoint of $X$ on an $n$ dimensional subspace. Given the form of $D_F$, it is obviously traceless. To show that the eigenvalues are symmetric with respect to zero, we use that for $A,B,C,D \in M^{n\times n}(\mathbb{C)}$ with $A$ invertible, and $AC=CA$ we have 
\begin{equation} \det  \begin{pmatrix}
A & B \\
C & D
\end{pmatrix} = \det (A\cdot D - C \cdot B) \:, \end{equation}
making it possible to calculate the characteristic polynomial for $D_F$,
\begin{align*}
    \chi_{D_F}(\lambda) &= \det (D_F- \lambda \mathbf{1}_{2n}) \\
    &= \det  \begin{pmatrix}
-\lambda \mathbf{1}_n & X \\
X^* & -\lambda \mathbf{1}_n
\end{pmatrix} = (-1)^n \det (X^* \cdot X - \lambda^2 \mathbf{1}_n)\\
&= (-1)^n \chi_{X^* \cdot X}(\lambda^2) \, .
\end{align*}
We conclude that $\chi_{D_F}(\lambda)=0$ implies that $\chi_{D_F}(-\lambda)=0$ and hence the eigenvalues of $D_F$ are symmetric with respect to zero. 
\QED 
 
 Continuing with our construction, we then consider the embedding of $D_F$ into $\text{BL}(\H)$ through an extension by $\mathbf{0}$ on the complement; i.e., for any embedding $\xi_a$, we define $D_F[\xi_a]:= D_F \oplus \mathbf{0} $ to act on $\xi_a(\H_F)\oplus(\xi_a(\H_F))^c= \H$. Given Lemma~\ref{lem:nn}, it is straightforward to see that $D_F[\xi_a]\in \F_n$. 

We can now proceed to construct the essential structures of which we learned from CFS that they allow for the generalization beyond classical continuum spacetimes. For clarity, here we choose to label the spacetime points and the corresponding physical wave functions by the index of the embedding  $\xi_a$
\begin{equation} \psi^{u_i}(a)= \pi_a u_i \:, \end{equation}
where $\pi_a$ is the projection operator on $\H$ with the embedding subspace $\xi_a(\H_F)$ as image. Furthermore, we can define the indefinite inner product by
\begin{equation} \Sl u | v \Sr_{D_F[\xi_a]} = \la u | D_F[\xi_a] v \ra_\H \qquad \text{(for all $u,v \in \xi_a(\H_F)$)}\:. \end{equation}
This allows us to define the generalized two-point correlator as above,
\begin{equation} P(a,b) := \sum_{i = 1}^N |{\psi^{u_i}(a)}\Sr_{D_F[\xi_a]} \Sl\psi^{u_i}(b)| \:. \end{equation}
Now it is clear that the setup of CFS is more general than what we can obtain from the above embedding of the local NCG structure into the global CFS structure. Because all the operators in the set
\begin{equation} \mathcal{D}_f:=\big\{D_F \,| \,D_F=D_F^* \,\text{ and  }  \,\{\sigma,D_F\}  =0 \big\} \end{equation}
have full rank, are traceless, and their eigenvalues are symmetric with respect to zero. While the first restriction has an equivalence in CFS when we restrict attention to regular spacetime points $\F_{\text{reg}}$, generic elements in this set are neither traceless, nor are their eigenvalues symmetric with respect to zero. Furthermore, it is interesting that the local trace $\tr_{\H_F}(D_F)$ does play a role in CFS, however, it is demoted to the role of a constraint in the causal action principle. An elementary result from CFS shows that the local trace has to be constant, $\tr(x)|_M=\text{const.}$, on the support of any minimizing measure, $\tr(D_F)=0$ of course being a special case of this. The action in CFS, even on the diagonal, is more general than just the trace of the local correlation operator, as discussed above.  
It would be an interesting question for future research whether measures defined on $\tilde{\xi}(\mathcal{D}_f)$ have any special properties with respect to the causal action. 

Now for GTD the situation is slightly less clear cut, but the key lesson from CFS seems to be applicable nonetheless. 
Given the fact that the dynamical variables $({\bf q}, {\bf p})$ of the ``atoms of spacetime'' in GTD are finite-dimensional matrices, they come with a natural $n$-dimensional Hilbert space $\H_{pq}$. We can now, of course, play the same game as with NCG and look at a larger, fundamental Hilbert space $\H$ and all possible embeddings of $\H_{pq}$ into $\H$. 
However, the question is: What should take the place of $D_F$ for GTD? As outlined in section 2.3.3 the operator $q_B = q_{BL} + q_{BR}$ adds internal contributions to the momentum operator, accounting for interacting bosonic fields. It is these terms that play the  natural analog of vector potentials $A$. In particular, when expanding the Lagrangian one obtains terms of the form $q_B^2$, which yield gauge kinetic terms, fermionic kinetic terms of the form $q_Bq_F$, and terms of the form $q_F^2$ which encode Higgs  potential terms. 
The Higgs  is a composite of the very fermions to which it gives mass. The finite internal part of the Dirac operator in NCG encodes Higgs fields. In GTD, Higgs fields arise from bilinear fermion terms of the form $q_F^2$ which occur when expanding out the Lagrangian. The Higgs also receives (kinetic) contribution from the scalar part of the octonions \cite{Raj:2022mcr}. As such there appears to be a direct correspondence between the data encoded in $D_F$ in the NCG approach to the standard model and that encoded in $q_F$ in the GTD approach~\cite{kaushik_2024}.

A key element of complexity added here is that, when we consider the evolution of a spacetime atom in $\H$, we have both the internal dynamics of $({\bf q}, {\bf p})$ as well as the potential dynamics of the embedding of each spacetime atom to consider. As an upshot of this additional flexibility, we would in turn gain access to the tools of correlation geometry provided by CFS. Since entanglement corresponds to a specific correlation pattern, this construction might give rise to potentially non-trivial ``entanglement'' between different spacetime atoms. This entanglement is a-priori dynamical. However, if the different spacetime atoms initially embed into disjoint subspaces, then for there to be an emergence of entanglement between the different spacetime atoms requires for the embeddings to be dynamical. Alternatively, one can fix which atoms are correlated at all to begin with (i.e., overlapping of the embedding spaces) and then just look at how the entanglement evolves purely based on the internal dynamics.

\section{Conclusion}
In the present paper, we compared three different approaches to fundamental physics, causal fermion systems, non-commutative geometry, and generalized trace dynamics, which all have their merits. If we look at the one domain where they all succeed, constructing a particle model that contains the standard model, then we draw one key lesson that we all believe to hold up in a successful theory of unification: The classical structure that one obtains in a suitable limit from a fundamental unifying theory will be a fiber bundle, and not the bare spacetime.

Furthermore, we proposed a general recipe on how the research program for NCG and CFS could be fused together. The key obstacle at this point being the development of efficient computational tools to check which constructions of NCG internal geometries are compatible with the causal action principle in the sense that they give rise to a minimizer.

For GTD finally we showed that the causal Lagrangian reduces to a trace on the diagonal, i.e., in the limit of non-interacting atoms of spacetime. This raises the question whether the current setup of the theory is sufficient to obtain a classical spacetime through entanglement of different atoms of spacetime. 

\Thanks{{{\em{Acknowledgments:}} 
We are grateful to the ``Universit\"atsstiftung Hans Vielberth'' for generous support.}

\bibliographystyle{amsplain}
\bibliography{tracedynamics}

\end{document}